\documentclass[11pt,a4paper,reqno]{amsart}

\usepackage[american]{babel}

\usepackage{amsmath,amssymb,graphics,epsfig,color,enumerate,psfrag}
\usepackage{dsfont}  

\usepackage{verbatim}
\newtheorem{Theorem}{Theorem}[section]

\newtheorem{TheoremA}{Theorem}

\newtheorem{Proposition}[Theorem]{Proposition}

\newtheorem{Remark}[Theorem]{Remark}

\newtheorem{Warning}[Theorem]{Warning}

\definecolor{refkey}{gray}{.75}
\definecolor{labelkey}{gray}{.5}

 \definecolor{darkgreen}{rgb}{0,0.4,0}

\definecolor{light}{gray}{.9}

\newcommand{\cA}{\ensuremath{\mathcal A}}

\newcommand{\cC}{\ensuremath{\mathcal C}}

\newcommand{\cE}{\ensuremath{\mathcal E}}

\newcommand{\cG}{\ensuremath{\mathcal G}}

\newcommand{\cI}{\ensuremath{\mathcal I}}

\newcommand{\bbC}{{\ensuremath{\mathbb C}} }

\newcommand{\bbE}{{\ensuremath{\mathbb E}} }

\newcommand{\bbP}{{\ensuremath{\mathbb P}} }

\newcommand{\bbR}{{\ensuremath{\mathbb R}} }

\newcommand{\bbZ}{{\ensuremath{\mathbb Z}} }

\newtheorem{CriterionA}{Criterion}

    \let\d=\delta  
\let\f=\varphi \let\g=\gamma       \let\l=\lambda
   \let\n=\nu         
  \let\s=\sigma \let\t=\tau   \let\th=\vartheta
  \let\z=\zeta
\let\D=\Delta   \let\G=\Gamma  \let\L=\Lambda

\author[A.\ Faggionato]{Alessandra Faggionato}
\address{Alessandra Faggionato.
  Dipartimento di Matematica, Universit\`a di Roma `La Sapienza'
  P.le Aldo Moro 2, 00185 Roma, Italy}
\email{faggiona@mat.uniroma1.it}

\author[V. Silvestri]{Vittoria Silvestri}
\address{Vittoria Silvestri. 
Statistical Laboratory, Centre for Mathematical Sciences, University of Cambridge, Wilberforce Road, Cambridge CB3 0WA, UK. }
\email{V.Silvestri@maths.cam.ac.uk}

\begin{document}

\begin{abstract}  Motivated by discrete kinetic models for non--cooperative molecular motors on periodic tracks, we consider random walks (also not Markov) on quasi one dimensional  (1d) lattices, obtained by gluing several copies of a fundamental graph in a linear fashion.
We show that, for a suitable class of quasi--1d lattices, the large deviation rate function associated to the position of the walker satisfies a Gallavotti--Cohen symmetry for any choice of the dynamical parameters defining the stochastic walk. This class includes the  linear model considered in \cite{LLM1}.  We also derive fluctuation theorems for the time--integrated cycle currents and discuss how the matrix approach of \cite{LLM1} can be extended to derive the above Gallavotti--Cohen symmetry for any Markov random walk on $\bbZ$ with periodic jump rates. Finally, we review in the present context some large deviation results of \cite{FS1}  and give some specific  examples with explicit computations. 


\medskip


\end{abstract}

\title[Fluctuation theorems for discrete kinetic models of molecular motors]{Fluctuation theorems for discrete kinetic models of molecular motors}

\maketitle

\section{Introduction}
Molecular motors  are special proteins able to convert chemical energy coming from ATP--hydrolysis into mechanical work,  allowing numerous physiological processes such as  cargo transport inside the cell, cell division, muscle contraction \cite{H}.  They are able to produce directed transport in an environment in which the fluctuations due to thermal noise are significant, achieving nonetheless an efficiency even higher than the one of macroscopic motors. 
In addition
synthetic molecular motors  have been  obtained and their improvements are under continuous investigation \cite{FY}.


Molecular motors haven been extensively studied both theoretically and  experimentally   (cf. \cite{JAP, KF3,Re,Ri,S} and references therein). We focus here on the large class of molecular motors (e.g. conventional kinesin) which work non--cooperatively and move along cytoskeletal filaments \cite{H}. Keeping in mind the  polymeric structure of these filaments, two main models have been proposed.  In the so called \emph{Brownian ratchet} model \cite{JAP,Re} the dynamics of the molecular motor is given by a one--dimensional diffusion in a  spatially  periodic potential  randomly switching its shape (indeed, along its mechanochemical cycle the molecular motor can be strongly or weakly bound to the filament, thus leading to a change in the interaction potential). 
The other paradigm  \cite{FK1,FK2,K1,K2,KF1,KF2,KF3,TF}, on which we concentrate here, is given by \emph{continuous time random walks}  (CTRW), along with a quasi one dimensional (quasi--1d) lattice  obtained by gluing several copies of a fundamental graph in a linear fashion.  CTRWs are thought  in the Montroll--Weiss sense \cite{MW}, and are also known as  \emph{semi--Markov processes satisfying the condition of direction--time independence} in the physical literature \cite{WQ}, as well \emph{Markov renewal processes} in the mathematical one \cite{As}.

\begin{figure}[!ht]
    \begin{center}
     \centering
  \mbox{\hbox{
  \includegraphics[width=0.8\textwidth]{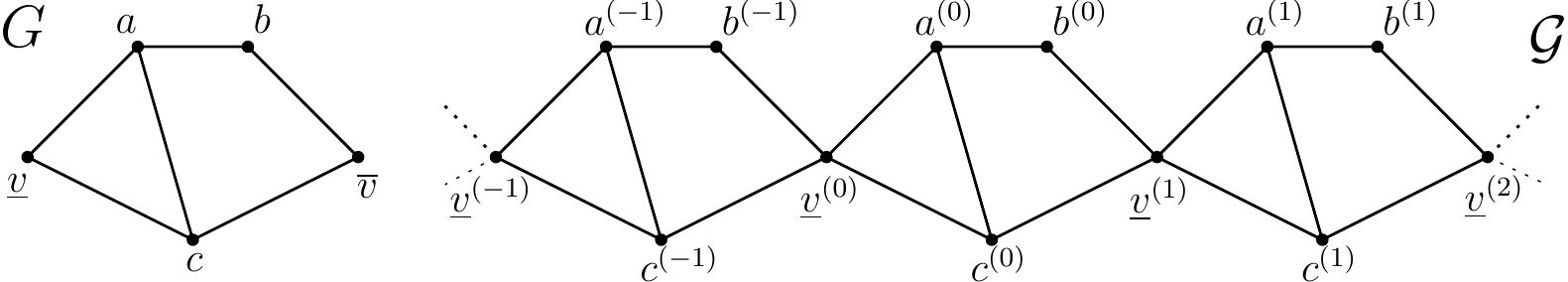}}}
            \end{center}
            \caption{The fundamental  graph $G$ with marked vertices $\underline{v}, \overline{v}$ (left)  the  quasi--1d lattice $\cG$ (right).} \label{pocoyo100}  
\end{figure}

The  above fundamental graph used to build a quasi--1d lattice is a  finite connected  graph $G$ with two marked vertices $\underline{v}$ and $\overline{v}$ (see Fig.\ \ref{pocoyo100}, left). For simplicity we assume that  $G$ has no  multiple edges or self--loops.   The associated quasi--1d lattice $\cG$  is then obtained by gluing several copies of $G$,  identifying the $\overline{v}$--vertex of one copy to  the $\underline{v}$--vertex  of the next copy (see Fig.\ \ref{pocoyo100}, right).  Given a vertex $v$ in $G$ and $n \in \bbZ$, we write $v^{(n)}
$ for the corresponding vertex in the $n$--th copy of $G$ in $\cG$. Since  $\overline{v}^{(n-1)}=\underline{v} ^{(n)}$, to simplify the notation we denote such a vertex by  $\underline{v} ^{(n)}$ throughout. Each site  $\underline{v}^{(n)}$ corresponds to a spot in the $n^{th}$ monomer of the polymeric filament to which the molecular motor can bind. The other vertices $v^{(n)}$ describe   intermediate conformational states that  the molecular motor achieves by conformational transformations, modeled by jumps along edges in $\cG$.  Note the periodicity of  the quasi--1d lattice $\cG$.

The evolution of the molecular motor is described by a CTRW  $(X_t)_{t \geq 0}$, taking values in the vertex set of the quasi--1d lattice $\cG$.
Once arrived at  vertex $x$, $X_t$ waits a random time with distribution  $\psi_x $ (that we assume to have finite mean) and then jumps to a neighboring vertex $y$ in $\cG$   with probability $p(x,y)>0$.    We assume that $\psi_x$ and $p(x,y)$ exhibit the same   periodicity of $\cG$. In what follows, we  call \emph{dynamical characteristics}  the above data $\psi_x$, $p(x,y)$. 

\begin{Warning}\label{ciocco} In the degenerate case that  $\psi_x$ is a delta measure, e.g. $\psi_x$ equals $\d_1$, the above CTRW reduces to the  so--called discrete time random walk. We do not restrict to distributions $\psi_x$ having  a probability density w.r.t. the Lebesgue measure, so that   $\psi_x$ can be composed by some delta measure as well.
\end{Warning}

We remark that when $\psi_x$ is the exponential distribution with mean $1/\l(x)$, then the resulting CTRW  is Markov and its density distribution $P_x(t):= P(X_t=x)$  satisfies the Fokker--Planck equation
  \begin{equation}\label{FP} 
  \frac{d}{dt} P_x(t) =\sum_{y} r(y,x) P_y(t)  -\sum _{y} r(x,y) P_x(t)\,, \qquad r(a,b):= \l(a) p(a,b) \,.
  \end{equation}
  In what follows, we assume that the random walk starts at $\underline{v}^{(0)}$, i.e. $X_0=\underline{v}^{(0)}$.

%
  
\begin{figure}[!ht]
    \begin{center}
     \centering
  \mbox{\hbox{
  \includegraphics[width=\textwidth]{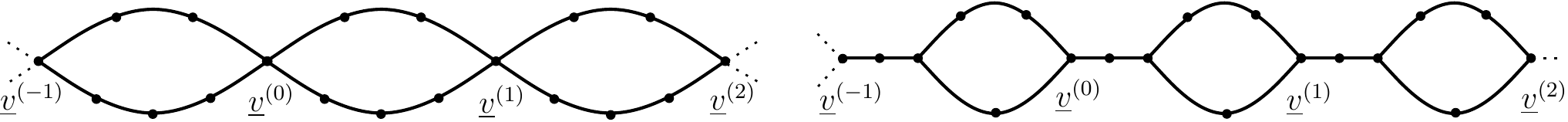}}}
            \end{center}
            \caption{Parallel chains model (left),   divided--pathway model (right).} \label{par_div_fig}
  \end{figure}

  As observed in \cite{TF}, the above formalism allows us to treat at once   several specific examples analyzed in the literature. For example, when the fundamental graph is given by a finite linear chain  with $N$ vertices,  we recover a CTRW on $\bbZ$  with  nearest--neighbor jumps and $N$--periodic dynamical characteristics 
  \cite{D,FK1,FK2}.  
   Supported by experimental results, CTRWs on more complex quasi--1d lattices have been considered in the biophysics literature \cite{DK,K1} (see  Fig.\ \ref{par_div_fig} for two examples).
 
Calling $V$ the set of vertices of the fundamental graph $G$, for $n \in \bbZ$ we define the \emph{$n^{th}$ cell}  as the set of vertices in $\cG$ of the form $v^{(n)}$ with $ v \in V \setminus \{\overline{v}\}$ (for example,  in Fig.\ \ref{pocoyo100} the $0^{th}$ cell is given by $\{ a^{(0)}, b^{(0)},c^{(0)}, \underline{v}^{(0)}\}$). Our aim is to investigate large fluctuations and associated symmetries of the \emph{cell process} $(N_t)_{t \geq 0}$, defined as $N_t=n$ if $X_t$ belongs to the $n^{th}$ cell, i.e. if $X_t=v^{(n)}$ for some $v \in V \setminus \{ \overline{v} \}$.  Trivially, the cell process determines the position of the molecular motor along the filament apart from an error of the same order of  the monomer size, which is negligible when analyizing  velocity, Gaussian fluctuations and large deviations.

  As shown in \cite{FS}, the cell process admits a  limit velocity 
$v_{\text{lim}}$ (i.e. $N_t /t \to v_{\text{lim}}$ almost surely) and has Gaussian fluctuations. A  large deviation principle  is  proved in  \cite{FS1}  (see Section \ref{panettone} for more details). We call $I: \bbR \to [0,+\infty]$ the associated large deviation function:
\begin{equation}\label{I_rate}
\bbP \left( N_t  \approx \vartheta t \right) \sim e^{-I(\vartheta) t  }\,, \qquad t \gg 1\,.
\end{equation}
In the last decades  some general principles, called \emph{fluctuation theorems}   and common to out--of--equilibrium systems,  have been formulated and intensively studied first for dynamical systems and then also  for stochastic processes  (see for example \cite{AG2,AG3,BFG2,CG,ES,GC,Ku,LS,SPWS}). For stochastic systems, they often correspond to  relations of the form  $J(\vartheta)= J(-\vartheta)- c \vartheta$, or similar, $c$ being a constant and $J$ being the rate function of an observable changing sign under  time inversion.  These last  relations are also called \emph{Gallavotti--Cohen type symmetries}, shortly  GC symmetries in what follows.  
Fluctuations theorems  have also been  investigated for small systems such as molecular motors \cite{AG1,FD,FS1, LLM1,LM1,LM2,LLM2,S}, and GC symmetries (in particular, for the velocity)  have been obtained for some special models. 
In particular, in \cite{LLM1,LM1,LLM2},  the authors derive a GC symmetry for the rate function of the velocity of a molecular motor described by a generic  Markov CTRW on $\bbZ$   with nearest--neighbor jumps and dynamical characteristics of periodicity two, which corresponds to \eqref{FP} with $r(a,b)$ of the  following form: $r(a,a\pm 1)= \xi_\pm$ if $a$ is even and $r(a,a\pm 1)= \z_\pm$ if $a$ is odd, for generic constants $\xi_\pm, \z_\pm >0$. This  GC symmetry for the velocity reads
\begin{equation}\label{gici} 
I(\vartheta)= I(-\vartheta)- c \vartheta \,, \qquad \vartheta \in \bbR\,,
\end{equation}   $I$ being the rate function of the cell process modulo rescaling by the length of  monomers in the polymeric filaments. For the above 2-periodic Markov CTRW it holds $c=\frac{1}{2} \ln \frac{ \xi _+ \z_+ }{ \xi_- \z_-}$.

\begin{figure}[!ht]
    \begin{center}
     \centering
  \mbox{\hbox{
  \includegraphics[width=0.55\textwidth]{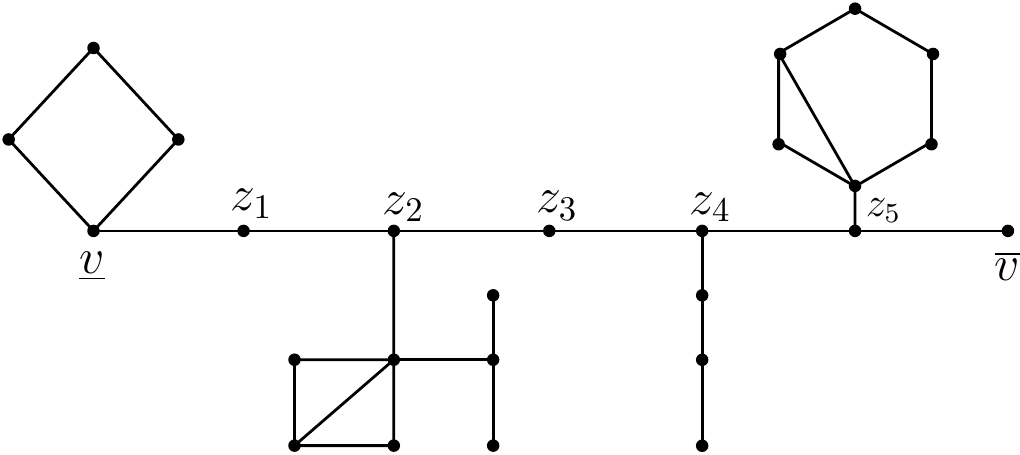}}}
            \end{center}
            \caption{Example of  a  \emph{$(\underline{v}, \overline{v})$--minimal}  graph $G$ with $\g= (\underline{v},z_1, \dots,  z_5, \overline{v} )$.}\label{sinti}
  \end{figure}

Since the above   CTRW with period $2$  is a simplified  model for the motion of  real molecular motors, a natural question concerns the validity of \eqref{gici} for a larger class of CTRWs, or even for all possible CTRWs  on quasi--1d lattices. For Markov CTRWs  we have shown in \cite{FS1} that \eqref{gici} is not universal, and in fact  \eqref{gici} is only universal in  the subclass of 1d lattices whose fundamental graph $G$  is   $(\underline{v}, \overline{v})$--minimal in the following sense: there exists  a unique self--avoiding path $\gamma$ in $G$  from $\underline{v}$ to  $ \overline{v}$. 
An example of   $(\underline{v}, \overline{v})$--minimal graphs $G$ is given in Fig.\ \ref{sinti}. 
Note that the graphs $G$ associated to the quasi--1d lattices in Fig.\  \ref{par_div_fig} are not  $(\underline{v}, \overline{v})$--minimal.

  We can now recall the characterization provided in \cite{FS1}:
  
   \begin{TheoremA}[\cite{FS1}]\label{lego_chima} 
 Suppose  that $X_t$ is a \underline{Markov} CTRW  on the quasi--1d lattice $\cG$, in particular it has   exponentially distributed waiting times and    transition rates $r(\cdot, \cdot)$ as in \eqref{FP}.  Then the following holds:
  \begin{itemize}
 \item[(i)] If $G$ is \emph{$(\underline{v}, \overline{v})$--minimal}, then  the cell process   $N_t$  satisfies the GC  symmetry 
\begin{equation}\label{verabila}
I(\th)=I(-\th)-\D \,\th\,,\qquad \forall \th\in \bbR\,,
\end{equation}
where\begin{equation}\label{alexey}
 \D= \ln \frac{ r(z_0,  z_1) r(z_1, z_2) \cdots r(z_{n-1},z_n ) }{  r(z_1,z_0) r(z_2,z_1) \cdots r(z_n,z_{n-1} ) }
\end{equation}
and $(z_0,z_1,z_2, \dots, z_{n-1} ,z_n ) $, with $z_0 = \underline{v}$ and $z_n= \overline{v}$, is the unique self--avoiding path from $\underline{v}$ to $\overline{v}$ in  $G$.  

\item[(ii)]
Vice versa,  if $G$  is not \emph{$(\underline{v}, \overline{v})$}--minimal,  then  the set of transition rates $r(\cdot, \cdot)$   for which the  GC symmetry \eqref{verabila} holds  for some $\D$ (which can depend on $r(\cdot, \cdot)$) has zero Lebesgue measure.

\end{itemize}
\end{TheoremA}

It is simple to verify (see Section \ref{colomba})  that the GC symmetry \eqref{gici} can be satisfied for very   special choices of the jump rates when  $G$ is not  \emph{$(\underline{v}, \overline{v})$}--minimal. In this case, due to the   above theorem, a   small perturbation of these rates typically  breaks the GC symmetry.

We point out that in \cite{FS} the GC symmetry for the LD rate function of the cell process is analyzed for a larger class of random processes, having a suitable regenerative structure. Moreover, it has been proved (cf. Theorems 4 and  8 in \cite{FS}) that the GC symmetry \eqref{gici} holds if and only if $X_{S_1}$ and $S_1$ are independent, where the random time $S_1$ is defined as  $S_1 := \inf \left\{ t \geq 0 \,:\, X_t  \in \{ \underline{v}^{(-1)} , \underline{v}^{(1)}\right \} $. For the Markov  random walk on a linear chain this independence had been pointed out  already in  \cite{DB} (see Remark \ref{natale100}).

\medskip

We also point out the above Theorem \ref{lego_chima}  is related to the theorem on page 584 of \cite{AGhor} (see also the discussion on cycle currents in Section \ref{GC_extra_cicli}). On the other hand, in the derivation of the equivalence stated in that theorem,  some additional arguments are necessary to get the difficult implication.

\medskip

The aim of the present work is the following:  (a) extend Theorem \ref{lego_chima}--(i) to generic  CTRWs (i.e. non Markov)  and give some sufficient condition assuring the GC symmetry \eqref{gici} for  non $(\underline{v}, \overline{v})$--minimal fundamental graphs (see Section \ref{sole1}), (b) derive fluctuation theorems for time--integrated cycle currents    in the case of generic CTRWs and  $(\underline{v}, \overline{v})$--minimal fundamental  graphs and, as a consequence, recover the GC symmetry \eqref{gici} independently from \cite{FS1} (see Section \ref{sole2}), (c) extend  the matrix approach outlined in \cite{LLM1}  to Markov CRTWs on general linear chain models,  getting also  the GC symmetry \eqref{verabila} (see Section \ref{sole3}), (d) give a short presentation of some results of \cite{FS1} in a less sophisticated language (see Section \ref{panettone}), (e) give specific examples with explicit computations (see Sections  \ref{colomba},  \ref{tagada}, \ref{RW_Z_exp} and \ref{RW_Z_gamma}).

\section{Main results}\label{sec_results}
In this section we present our main results, postponing their derivation to the next sections and to the appendixes.

\subsection{Extension of Theorem \ref{lego_chima}--(i) to generic CTRWs}\label{sole1}
We consider generic CTRWs on $\cG$, i.e. also non Markov.  As a first result we give a sufficient condition assuring that the GC symmetry \eqref{gici} holds for some constant $\D$ (for a sufficient and necessary condition see Criterion  \ref{criceto} in Appendix
\ref{castelluccio_proof}). This condition is trivially satisfied in $(\underline{v}, \overline{v})$--minimal graphs $\cG$,  thus leading 
 to the extension of  Theorem \ref{lego_chima}--(i)  to non Markov CTRWs.

\begin{TheoremA} \label{castelluccio} Consider a generic CTRW
  $(X_t) _{t \geq 0}$  on the quasi--1d lattice $\cG$ with dynamical characteristics $p(x,y)$ and $\psi_x$. Then   the cell process   $N_t$  satisfies
the GC symmetry \eqref{verabila} for some constant $\D$ if 
\begin{equation}
\prod_{i=0}^{m-1} p(x_i, x_{i+1})= e^\D \prod_{i=0}^{m-1}p(x_{i+1}, x_{i})
\end{equation}
for any self--avoiding path $(x_0, x_1, \dots, x_m)$ from $\underline{v}$ to $\overline{v}$ in the fundamental graph $G$  ($x_0=\underline{v}$, $x_m=\overline{v}$).

As a consequence, 
if $G$ is \emph{$(\underline{v}, \overline{v})$--minimal}, then  the cell process   $N_t$  satisfies the GC  symmetry \eqref{verabila} 
where now \begin{equation}\label{alexey_biz}
 \D= \ln \frac{ p(z_0,  z_1) p(z_1, z_2) \cdots p(z_{n-1},z_n ) }{  p(z_1,z_0) p(z_2,z_1) \cdots p(z_n,z_{n-1} ) }
\end{equation}
and $(z_0,z_1,z_2, \dots, z_{n-1} ,z_n ) $, with $z_0 = \underline{v}$ and $z_n= \overline{v}$, is the unique self--avoiding path from $\underline{v}$ to $\overline{v}$ in  $G$.  
\end{TheoremA}
Note that for Markov CTRWs expressions \eqref{alexey} and \eqref{alexey_biz} indeed coincide.

\smallskip

The theorem is a immediate consequence of Criterion \ref{criceto} discussed in Appendix \ref{castelluccio_proof}. 

\smallskip 

\begin{Remark}
When considering discrete time RWs (recall Warning \ref{ciocco}) it is possible to exhibit examples of fundamental graphs $G$ which are not $(\underline{v}, \overline{v})$--minimal  and such that the GC symmetry \eqref{gici} holds for any choice of the jump probabilities $p(x,y)$. We refer to Section \ref{tagada} for an example.
 \end{Remark}

\subsection{GC symmetries for cycle currents}\label{sole2}
As next result we show that, for $(\underline{v}, \overline{v})$--minimal fundamental  graphs, the GC symmetry \eqref{gici} is indeed a special case  of a fluctuation theorem for 
cycle currents  (see e.g.\   \cite{AG2,AG3,F,FD}).  As a consequence  we give, among others,  an alternative derivation of \eqref{gici} for $(\underline{v}, \overline{v})$--minimal fundamental  graphs, which is based on cycle theory and does not use preliminary facts from \cite{FS1} as the above cited Criterion \ref{criceto}.

\begin{figure}[!ht]
    \begin{center}
     \centering
  \mbox{\hbox{
  \includegraphics[width=0.65\textwidth]{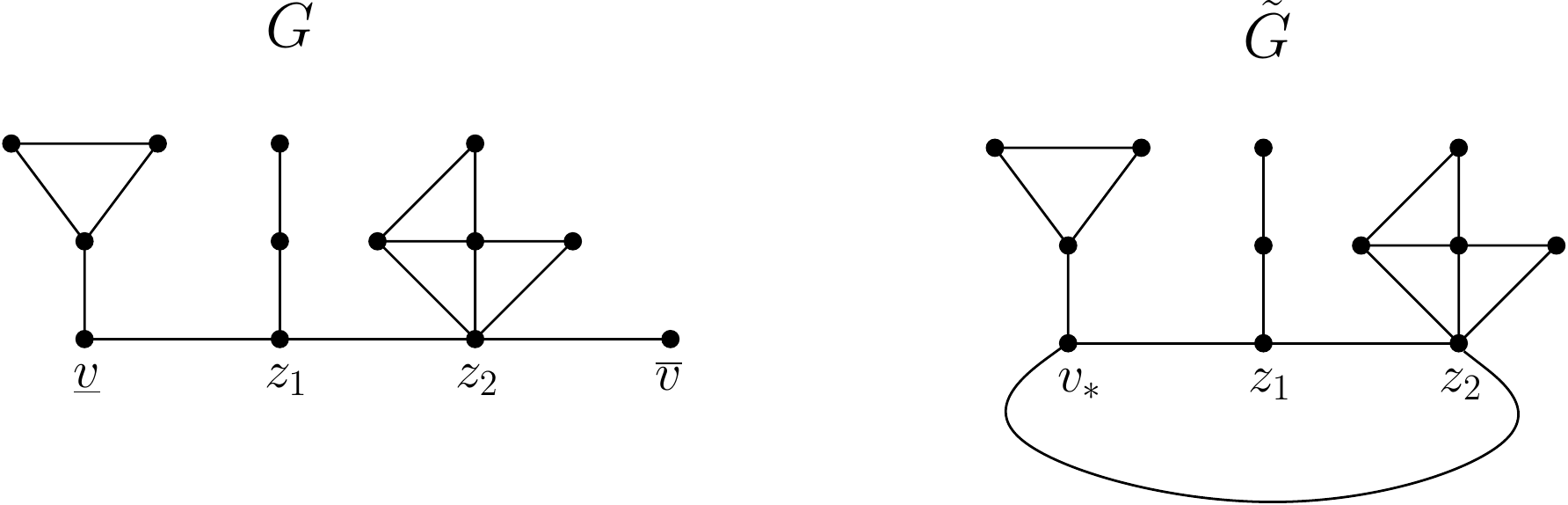}}}
            \end{center}
            \caption{The fundamental graph  $ G$ 
            and the associated graph $\tilde G$ obtained by gluing $\underline{v}$ and $\overline{v}$. }\label{fig:incollo}
  \end{figure}

We  present  here our result giving more  details and precise definitions in Section \ref{GC_extra_cicli}.
To this aim, we assume $G$ to be $(\underline{v}, \overline{v})$--minimal and 
 we denote by $\tilde G$
 the new finite graph  obtained from $G$ by gluing together $\underline{v}
$ and $\overline v$ in a single vertex called $v_*$  (see Fig.\ \ref{fig:incollo}).

We denote by $\cC_1$ the cycle in $\tilde G$ corresponding 
to  the unique self--avoiding path   $(z_0, z_1, \dots, z_n)$ from $\underline{v}$ to  $\overline{v}$ in $G$, and we  call $\cC_2, \dots, \cC_m$ the other cycles in $\tilde G$ which form, together with $\cC_1$, a cycle basis according to Schnackenberg's construction. We also define the affinity $\cA(\cC)$ of a cycle $\cC$ as 
  \begin{equation}\label{panda_grande}
 \cA(\cC): = \ln
   \prod _{i=0} ^{k-1} \frac{ p( x_i, x_{i+1})
   }{  p (x_{i+1},x_i) }  \qquad\text{ if }\:
 \cC= ( x_0,x_1, \dots, x_k)\,, \; \;x_0=x_k\,.
 \end{equation}

 Due to the periodicity of the dynamical characteristics, the CTRW $X_t$ naturally induces a CTRW  $Y_t$  on $\tilde{G}$. We then consider  the path in $\tilde G$ given  by  the  vertices   visited by 
  $Y$ up to time $t$ and complete it to get a cycle $\cC_t$ in $\tilde G$, e.g. by adding an extra path of minimal length ending at  the initial point.  Finally we decompose the random cycle $\cC_t$ in the above cycle basis: $\cC_t= \sum _{i=1}^m a_i(t) \cC_i$. The random coefficients $a_i(t)$'s are  also called \emph{time--integrated cycle currents}, and for them we derive in Appendix \ref{proof_teo_cicli}  the following fluctuation theorems:
\begin{TheoremA}\label{cicli} Suppose that 
 $G$   is $(\underline{v}, \overline{v})$--minimal and  let $(X_t)_{t\geq 0}$ be a  generic CTRW on the associated quasi--1d lattice $\cG$. 
Then the random vector $\frac{1}{t}(a_1(t)  , a_2(t), \dots , a_m(t))$ satisfies a LDP  with  speed $t$ and good\footnote{''Good'' means that the level sets of $\cI$ are compact} rate function $\cI$.  Calling $\cI(\th_1, \th_2, \dots, \th_m)$ the associated rate function,  roughly  we have 
\begin{equation}
\bbP\left[\frac{1}{t} (a_1(t), \dots, a_m(t) ) \approx ( \th_1, \dots,  \th _m) \right]\sim e^{ - t \cI(\th_1, \th_2, \dots, \th_m)}\,.
\end{equation} 
%
Moreover the following  GC symmetries hold:
\begin{align}
& \cI( \th_1, \th_2, \dots, \th_m) = \cI(-\th _1, -\th _2, \dots, -\th_m) -\sum _{i=1}^m \th _i \cA( \cC_i)\,, \label{crimea}\\
& \cI(\th_1, \th_2 \dots, \th_m)= \cI( \th_1, -\th_2, \dots, -\th_m) - \sum _{i=2}^m \th_i \cA(\cC_i)\,, \label{ucraina}\\
& \cI(\th_1, \th_2, \dots, \th_m)= \cI(-\th_1, \th_2, \dots, \th_m)-\th_1 \cA( \cC_1) \,. \label{mosca}
\end{align}
As a consequence,   the LD rate function $I(\vartheta)$ of the cell process introduced in \eqref{I_rate} fulfills  the GC symmetry
\begin{equation}\label{lorenza}
 I(\th)=I(-\th)- \cA (\cC_1)  \,\th\,,\qquad \forall \th\in \bbR\,.
 \end{equation}
\end{TheoremA}

%
%

Let us also remark that for Markov CTRWs the symmetry \eqref{lorenza} reduces  to \eqref{verabila}, since  $ \cA(\cC_1)= \D$.

For Markov CTRWs  \cite{FD}, but also for a larger class of  CTRWs,
one can show that the function 
\[ Q(\l_1, \l_2, \dots, \l_m) := \lim _{t \to \infty} - \frac{1}{t} \ln \bbE\left[ e^{-\sum _{i=1}^m \l_ia_i(t) }\right]\,, \qquad  (\l_1, \l_2, \dots, \l_m) \in \bbR^m
\]
is well posed and  it holds 
 \begin{equation}\label{roma}\cI(\th_1, \th_2, \dots, \th_m):= \sup _{ (\l_1,\l_2, \dots, \l_m)\in \bbR^m }\bigl\{ -\sum_{i=1}^m \th_i \l_i + Q(\l_1, \l_2, \dots, \l_m) \bigr\}\,,\end{equation}
Moreover,  via Legendre transform, the above identities \eqref{crimea}, \eqref{ucraina} and \eqref{mosca} correspond respectively to the following \eqref{lord1}, \eqref{lord2} and \eqref{lord3}:
\begin{align}
&  \label{lord1}
Q(\l_1 , \l_2 , \ldots , \l_m) = Q( \cA (\cC_1) - \l_1 , \cA (\cC_2) - \l_2 , \ldots , \cA (\cC_m) -\l_m )
\,,\\
& \label{lord2}
Q(\l_1 , \l_2 , \ldots , \l_m)  = Q(\l_1 , \cA (\cC_2) - \l_2 , \ldots , \cA (\cC_m) -\l_m )
\,, \\
& \label{lord3}
Q(\l_1 , \l_2 , \ldots , \l_m) = Q( \cA (\cC_1) - \l_1 , \l_2 , \ldots , \l_m )
\,.
\end{align}

\subsection{Derivation of the GC symmetry \eqref{verabila} for Markov CTRWs  on the linear chain by the matrix approach}\label{sole3}
When the CTRW on the quasi--1d lattice $\cG$ is Markov, then the  LD  rate function $I$ of the cell process $N_t$ can be expressed as the Legendre transform  of the maximal eigenvalue of a suitable  matrix depending by a scalar parameter. In \cite[Theorem 3]{FS} a general formula is derived by generalizing the matrix approach used in \cite{LLM1}.

\begin{figure}[!ht]
    \begin{center}
     \centering
  \mbox{\hbox{
  \includegraphics[width=0.7\textwidth]{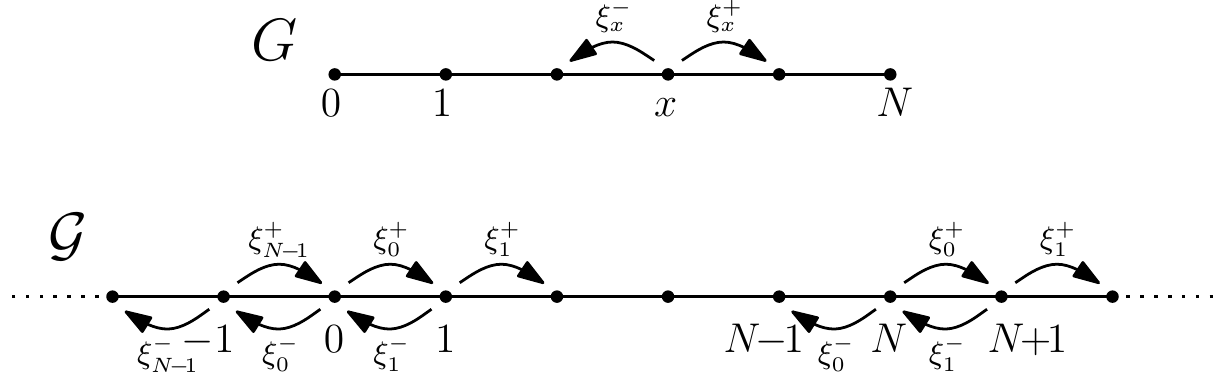}}}
         \end{center}
         \caption{The linear chain graph $G$ (up), and the associated quasi--1d lattice $\cG$ (down).} \label{linear_pic}
  \end{figure}

We restrict here to Markov CTRWs on a linear chain and show how one can derive 
 the GC symmetry \eqref{verabila} by the matrix approach. To make the discussion self--contained we briefly recall how to express the LD rate function in terms of the above maximal eigenvalue. To this aim 
let $G$ be the linear chain graph of Fig.\ \ref{linear_pic}, i.e. $G=(V,E)$ with $V=\{ 0,1,\ldots , N\}$, $E=\{ (x,x+1 ) , x=0 , \ldots , N-1\}$ and $\underline{v}=0 $, $\overline{v} =N$. If $\cG$ denotes the associated quasi--1d lattice, then $\cG$ can be identified with $\bbZ$ with periodic jump rates. We therefore take $\bbZ$ to be the vertex set of $\cG$, and denote 
 by $\xi_x^\pm$, $ x \in \bbZ$, the rate associated to the edge $(x,x\pm 1)$. Finally, set  $r(x) = \xi_x^- + \xi_x^+$. Then the Markov CTRW $X_t$ waits at $x$ an exponentially distributed  time of mean $1/r(x)$, and then jumps to either $x+1$ or $x-1$ with probability $\xi_x^+ /r(x)$ and $\xi_x^-/r(x)$ respectively.  Note that $\xi_x^\pm= \xi_{x+N}^\pm$ and $r(x)= r(x+N)$ for any $x \in \bbZ$ and that the constant $\D$ in \eqref{alexey} is now given by $\D= \ln \frac{ \xi_0^+ \xi_1^+\cdots \xi^+_{N-1} }{  \xi_0^- \xi_1^-\cdots \xi^-_{N-1}}
$.

Let us first consider the case $N \geq 3$. Given $\l \in \bbR$, we introduce the $N\times N$  matrix $\cA(\l)$, defined  as follows for $0\leq i,j \leq N-1$:
\begin{equation} \label{matrixA}
\cA (\l) _{i,j} = 
\begin{cases}
-r(i) & \text{ if } i=j\,,\\
\xi_{j} ^+ & \text{ if }0<i \leq N-1,\;\; j=i-1\,,\\
\xi_{j}^- & \text{ if } 0\leq i < N-1, \;\;j=i+1\,,\\
\xi_0^- e^{-\l} & \text{ if } i=N-1, j=0\,,\\
\xi_{N-1}^+ e^\l & \text{ if } i=0, j=N-1\,,\\
0 & \text{ otherwise}\,.
\end{cases}
\end{equation}
For example, for $N=3$ we have 
\[\cA (\l ) = 
	\left( 
	\begin{matrix}
	- r(0) & \xi_1^-  & \xi_2^+ e^\l  \\
	 \xi_0^+ & - r(1) & \xi_2^- \\
	 \xi_0^{-} e^{-\l} & \xi_1^+ & -r(2)
	\end{matrix} 
	\right)\,.
\]

Following the approach of \cite{LLM1} for the $2$--periodic linear model, we introduce the function 
\[ F(x,\l,t):= \sum _{k \in \bbZ} e^{\l k} \bbP (X_t= x+kN )= \bbE\bigl[ e^{\l N_t} \mathds{1}(X_t=x+N_t N ) \,\bigr] , \]
where, we recall, $N_t$ is the  cell number of $X_t$. 
By the Markov property of $X_t$  we have $\partial _tF(x,\l,t)= \xi_{x-1}^+   F(x-1,\l,t)+ \xi_{x+1}^- F(x+1,\l,t)- r(x) F(x,\l,t)$. Using that $F(-1,\l,t)= e^\l F(N-1,\l,t)$ and $F(N,\l,t)= e^{-\l} F(0,\l,t)$, we conclude that
\begin{equation}\label{jc}
\partial_t F(x, \l, t)= \sum _{0\leq y \leq N-1} \cA(\l)_{x,y} F(y,\l,t) \,, \qquad  0\leq x \leq N-1\,,
\end{equation}
and therefore $ F(x, \l, t)= \sum _{0\leq y \leq N-1} [e^{t \cA  (\l) }]_{x,y} F(y,\l,0)$.
When $N=2$, \eqref{jc} remains valid with $\cA(\l)$ defined as 
\[\cA (\l ) = 
	\left( 
	\begin{matrix}
	- r(0) &  \xi_1^+e^\l+\xi_1^{-}    \\
	\xi_0^++ \xi_0^- e^{-\l}  & -r(1) 
	\end{matrix} 
	\right)
\]
Since on the other hand $ \bbE( e^{\l N_t})= \sum _{0\leq x \leq N-1} F(x,\l,t)$, the  Perron--Frobenius theorem gives\footnote{$\Re(x)$ denotes the real part of the complex number $x$.}
\begin{equation}\label{telaviv2}
 \bbE( e^{\l N_t}) \approx e^{ t \L(\l) }   \,,   \qquad \L(\l):= 
\max \{ \Re (\g) \,:\, \g \text{ eigenvalue of } \cA (\l) \}\,. 
\end{equation}
By G\"artner--Ellis theorem, the cell process satisfies a LD principle with rate function $I$ given by
\begin{equation}\label{petardi}
 I(\th)=\sup_{ \l \in \bbR} \{ \th \l - \L(\l) \} \,, \qquad  \th \in \bbR \,.
\end{equation}
Having \eqref{petardi} the  GC symmetry \eqref{verabila} follows  from the equality
\begin{equation}\label{zoo}
\L(\l)= \L(-\D-\l)\,, \qquad \l \in \bbR \,,
\end{equation}
 with $\D$ defined according to \eqref{alexey}. This is in turn a consequence of the following result:
  \begin{Proposition}\label{suricato} Let $\D= \ln \frac{ \xi_0^+ \xi_1^+\cdots \xi^+_{N-1} }{  \xi_0^- \xi_1^-\cdots \xi^-_{N-1}}
$. Then 
 there exists an invertible  matrix $U$ such that  
 \begin{equation}\label{patate}
  U^{-1} \cA(\l) U = \cA ^{\rm T} (-\D-\l) \qquad \forall \l \in \bbC\,,
  \end{equation} 
  $\cA ^{\rm T} (-\D-\l)$ being the transpose of $\cA (-\D-\l)$.
   In particular,  for the linear chain graph identity \eqref{zoo} is satisfied as well as  the GC symmetry \eqref{verabila}.
 \end{Proposition}
It is known that any  square matrix $A$ is similar to its transpose $A^{\rm T}$ \cite{TZ}, i.e. $\exists$
an invertible matrix $U$ such that $U^{-1} A U =A^{\rm T}$. Hence,  once proved  \eqref{patate}, one immediately gets  that $\cA(\l)$ and $\cA(-\D-\l)$ have the same spectrum and therefore the conclusion of the proposition becomes trivial by the above discussion. 

\subsection{Further results}
Four  specific examples are discussed in Sections \ref{colomba},  \ref{tagada}, \ref{RW_Z_exp} and \ref{RW_Z_gamma}. We briefly comment on  them.
 The derivation of Theorem \ref{lego_chima}--(ii), given in \cite{FS1}, is mathematically involved. On the other hand,  in  Section \ref{colomba}  we consider 
 a parallel chains model (whose fundamental graph is  not $(\underline{v}, \overline{v})$--minimal) and show by direct computations that usually the GC symmetry \eqref{gici} is not satisfied. In particular, we recover in a specific example the content of Theorem \ref{lego_chima}--(ii).
In Section \ref{tagada}, by considering  discrete time RWs (recall Warning \ref{ciocco}),  we exhibit an example of fundamental graph $G$ which is  not $(\underline{v}, \overline{v})$--minimal  and such that the GC symmetry \eqref{gici} holds for any choice of the jump probabilities $p(x,y)$. 
   Finally, in Sections \ref{RW_Z_exp} and \ref{RW_Z_gamma} we consider spatially homogeneous CTRWs on $\bbZ$   with waiting times having respectively exponential and gamma distribution, and compute explicitly  several quantities related to large deviations introduced in Section \ref{panettone} (in particular, the LD rate function for the hitting times and the LD rate function for the cell process).
 
 \subsection{Outline of the paper}
 As already  pointed out,  a crucial feature of the  CTRWs on quasi--1d lattices is a regenerative structure (several results of \cite{FS1} are indeed valid for stochastic processes exhibiting such a regenerative structure,   not necessarily  CTRWs). We  explain this regenerative structure  in Section \ref{spa}.  In Section \ref{panettone} we recall the main results of \cite{FS1} applied to the present context, while in Section \ref{GC_extra_cicli} we recall some basic facts on cycle currents 
 and discuss in detail the objects involved in the cycle fluctuation theorems stated in Theorem   \ref{cicli}.
  Some of these results will be used in our proofs.
In Sections   \ref{colomba}, \ref{tagada}, \ref{RW_Z_exp} and \ref{RW_Z_gamma} we discuss the above mentioned example. Appendixes \ref{castelluccio_proof}, \ref{proof_teo_cicli} and 
\ref{GC_extra_matrix} will be devoted to the derivation of Theorem \ref{castelluccio}, Theorem \ref{cicli} and Proposition \ref{suricato} respectively. Finally, Appendix \ref{aiutino} contains some minor technical facts.

\section{Regenerative structure and skeleton process} \label{spa}  
\begin{figure}[!ht]
    \begin{center}
     \centering
  \mbox{\hbox{
  \includegraphics[width=0.7\textwidth]{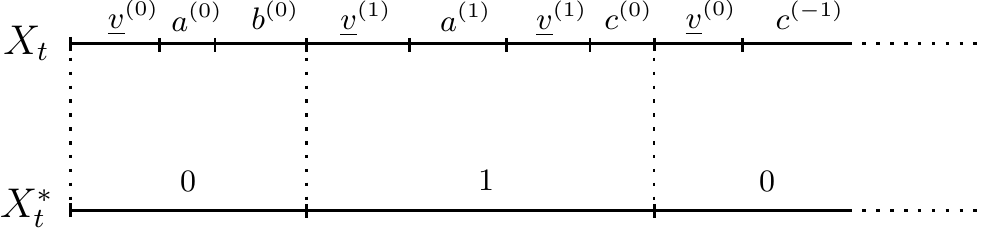}}}
         \end{center}
         \caption{Example of a trajectory  $(X_t)_{t \geq 0 }$ and the associated trajectory  $ (X_t^*)_{t \geq 0 }$ referred to the quasi--1d lattice $\cG$ of Fig.\ \ref{pocoyo100}.}  \label{esempietto_fig} 
\end{figure}

In this section we explain the regenerative structure behind the CTRWs on $\cG$.
To this aim we   introduce  a coarse--grained version of $X_t$,  called \emph{skeleton process} $ (X_t^*)_{t \geq 0 }$ with values in $\bbZ$. More precisely, we set $X^*_t =n$ if $\underline{v}^{(n)}$ is the last vertex of the form $\underline{v}^{(k)}$ visited by $(X_s)_{0\leq s  \leq t}$ (see the example in Fig.\ \ref{esempietto_fig}).  In the applications to molecular motors,  the skeleton process contains  all the relevant information, since it allows to determine the position of the molecular motor up to an error of the same order of the monomer size. 
 
   Note that $|N_t -X_t^*  | \leq 1$, and therefore the skeleton process and the cell process have the same asymptotic behaviour  and  large deviations. 
   
\begin{figure}[!ht]
    \begin{center}
     \centering
   \mbox{\hbox{
   \includegraphics[width=0.7\textwidth]{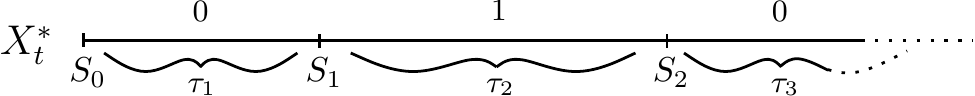}}}
         \end{center}
         \caption{Jump times $S_i$ and  inter--arrival
        times $\t_i$  for the trajectory of the skeleton process of 
         Fig.\ \ref{esempietto_fig}. Note that $w_1=+1$ and  $w_2=-1$.
 }\label{nepi}   
 \end{figure}

   The technical advantage of dealing with the skeleton process instead of the cell process  comes from  the following  regenerative structure.  
  Consider the sequence $S_1 < S_2 < \dots $ of jump times for the skeleton process $X^*_t$, set $S_0:=0$,  call $\t_i:= S_i- S_{i-1}$ the   inter--arrival times and $w_i:= X^*_{S_i} -X^*_{S_{i-1}} \in \{-1,+1\}$ the jumps of the skeleton process (see Fig.\ \ref{nepi}). By our assumptions on $X_t$, we get that the sequence $(w_i, \t_i)_{ i \geq 1}$ is given by \emph{ independent and identically distributed random vectors} and it fully characterizes the skeleton process itself. 
  
\section{Time integrated cycle currents and affinity}  \label{GC_extra_cicli}

In this section we restrict to $(\underline{v}, \overline{v})$--minimal fundamental graphs $G$ and 
  apply the cycle theory  (see e.g.   \cite{AG2,AG3,F,FD})  to formulate      fluctuation theorems 
for cycle currents also for non--Markovian CTRW (cf. Theorem \ref{cicli}).

We denote by  $\g=(z_0, z_1, \dots, z_n)$ the unique self--avoiding path from $\underline{v}$ to  $\overline{v}$ in $G$,  hence with  $z_0 = \underline{v}$, $z_n = \overline{v}$. We assume that $n \geq 3$ without loss of generality, since  
the cases $n=1,2$ can be reduced to the one above by doubling or tripling the fundamental cell, as  explained in Appendix \ref{aiutino} (see also Fig.\ \ref{aiutino_fig} therein). 

\begin{figure}[!ht]    \begin{center}
     \centering
  \mbox{\hbox{
  \includegraphics[width=0.65\textwidth]{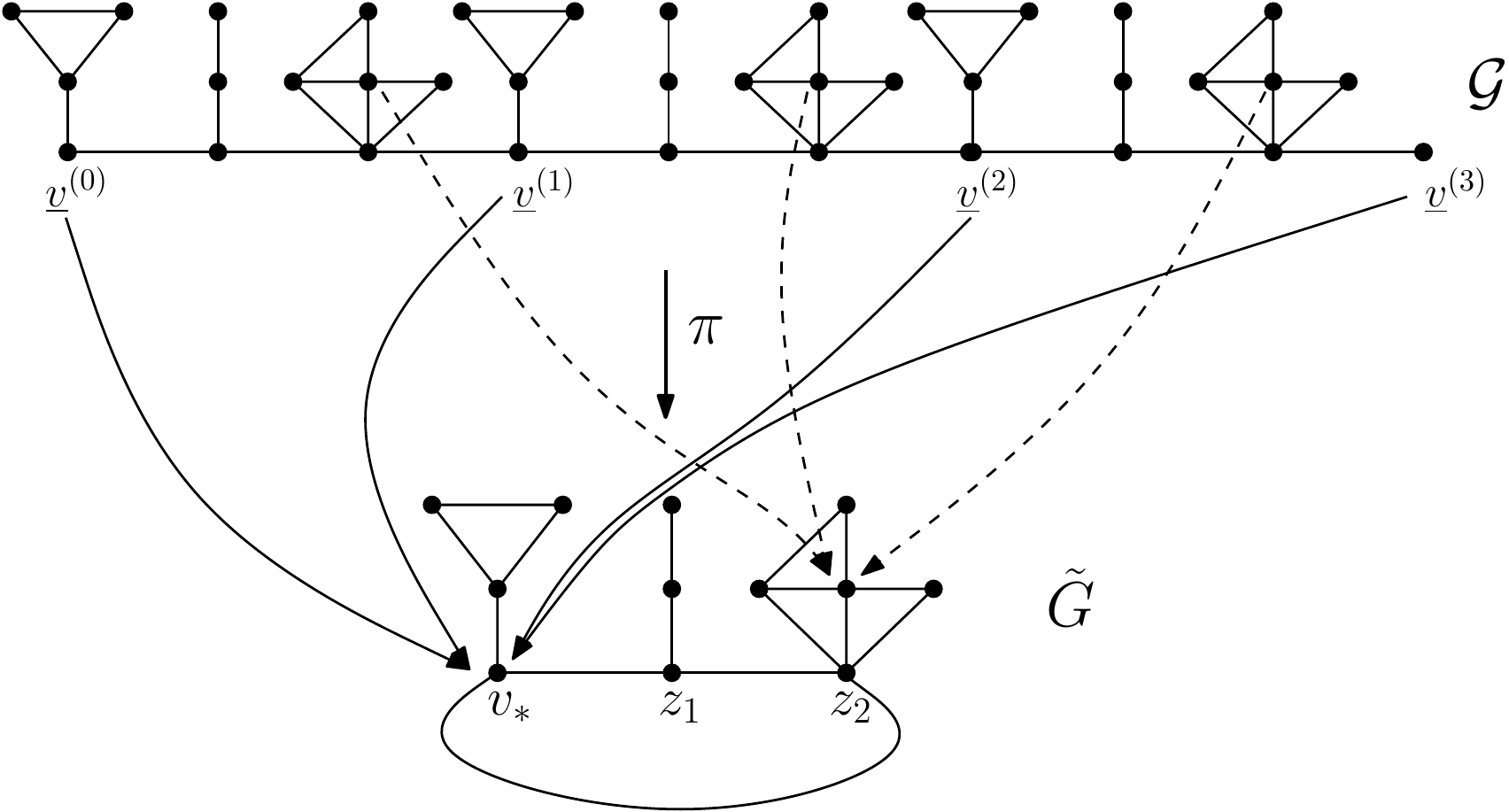}}}
            \end{center}
            \caption{The natural projection $\pi: \cG \to \tilde G$.}
            \label{fig:proiettore}
  \end{figure}

Let $\tilde G$ denote a new finite graph  obtained from $G$ by gluing together $\underline{v}
$ and $\overline v$ in a single vertex called $v_*$  (see Fig.\ \ref{fig:incollo}). 
We denote by $\pi: \cG \to \tilde G$   the natural graph projection (see Fig.\ \ref{fig:proiettore}) and introduce the projected process $Y_t:= \pi(X_t)$ having values in $\tilde{G}$. As explained in formula \eqref{sanpietro} below, one can recover the asymptotic behavior of the skeleton process $X_t^*$ (and therefore of the cell process $N_t$) by analyzing the currents of the  projected process $Y_t$.

%
  
%
%

Let us briefly recall some concepts from cycle theory (see e.g. \cite{AG2,BFG2,FD,Sc}). A cycle $\cC$ in $\tilde{G}$ is described by a path $(x_0, x_1, \dots, x_k)$ along edges of $\tilde{G}$ such that $x_0=x_k$. 
Given a cycle $\cC$ and  two neighboring vertices $x,y$ in  $\tilde G$, we define 
  $N_{x,y} (\cC)$ as the number of appearances of the string $(x,y)$ in $\cC$ minus 
 the  number of appearances of the string $(y,x)$ in $\cC$ (i.e. the number of jumps from $x$ to $y$ minus the number of jumps from $y$ to $x$ performed by  the cycle $\cC$).  
 We can make the cycle space into a real vector space by considering formal  linear combinations of cycles and using the identification 
 \begin{equation}
\label{ab}
 \sum _{i =1}^m a_i \cC_i =  \sum _{j =1}^k b_j \cC'_j
 \end{equation}
 whenever 
 $ \sum _{i =1}^m a_i N_{x,y}( \cC_i)  =  \sum _{j =1}^k b_j N_{x,y}(\cC'_j)$ for any neighboring vertices  $x,y$.

\begin{figure}[!ht]
    \begin{center}
     \centering
  \mbox{\hbox{
  \includegraphics[width=0.8\textwidth]{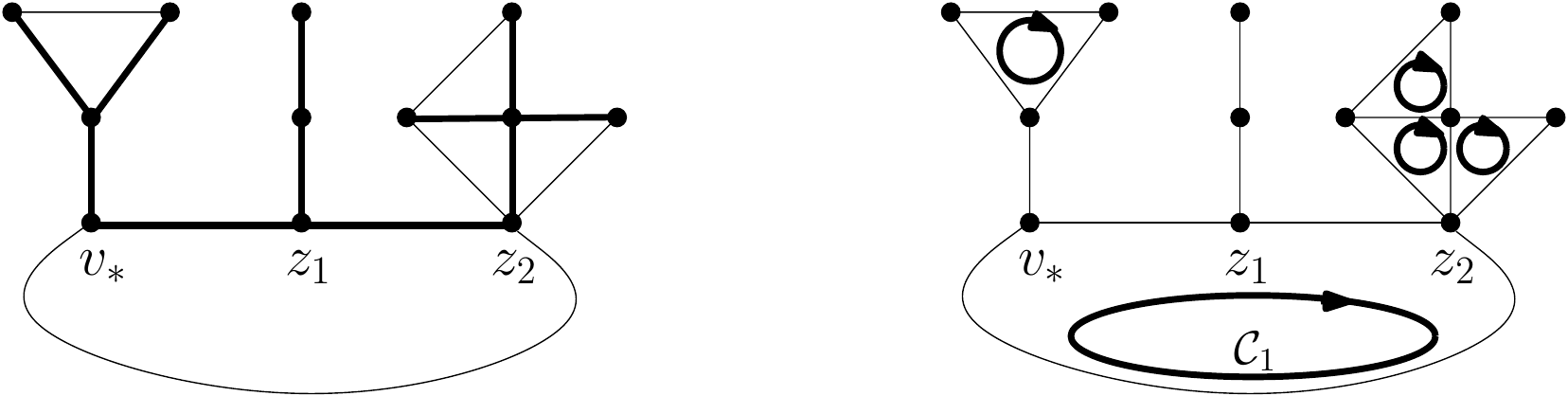}}}
            \end{center}
            \caption{Schnackenberg's construction of the cycle basis $\cC_1, \dots,\cC_m$. In the left picture, the spanning  tree is drawn in bold lines.}\label{fig:lumachina}
\end{figure}

To the path   $\g=(z_0, z_1, \dots, z_n)$ we associate the cycle $(v_*, z_1, z_2, \dots, z_{n-1}, v_*)$ in the graph $\tilde G$.  
%
%
%
Let us now fix a cycle basis  $\cC_1, \cC_2,  \dots, \cC_m$ in $\tilde G$  with  $\cC_1= (v_*, z_1, z_2, \dots, z_{n-1}, v_*)  $.   This can be done according to Schnackenberg's construction as follows. We take a spanning tree (i.e.\ a subgraph of $\tilde G$ without loops  which contains all vertices of $\tilde G$) containing the linear chain $(v_*, z_1,z_2, \dots, z_{n-1})$. Given an edge $\{x,y\}$ in $\tilde G$ not belonging to the spanning tree, there exists a unique self--avoiding cycle $\cC $  (apart from orientation and starting point) in the graph obtained by adding the edge $\{x,y\}$ to the spanning tree. Just take one $\cC$, fixing arbitrarly orientation and starting point. The collection of cycles obtained by varying the edge $\{x,y\}$ in this procedure forms a cycle basis. Note that this basis contains the cycle  $(v_*, z_1, z_2, \dots, z_{n-1}, v_*) $,  which is indeed associated to the edge $\{x,y\} = \{v_*, z_{n-1}\}$ (see Fig.\ \ref{fig:lumachina}). Note also that, since $G$ is $(\underline{v}, \overline{v})$--minimal,  
the cycle $\cC_1$ has no edge in common with $\cC_2, \dots, \cC_m$.

\begin{figure}[!ht]
    \begin{center}
     \centering
  \mbox{\hbox{
  \includegraphics[width=0.7\textwidth]{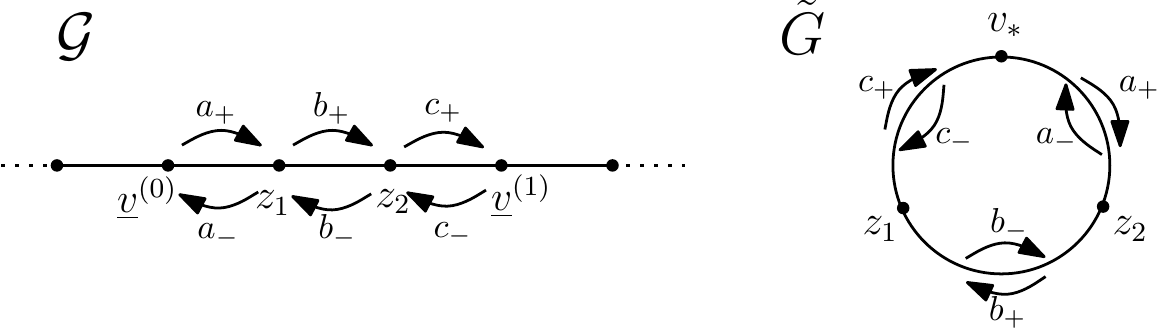}}}
            \end{center}
            \caption{The quasi--1d lattice $\cG$ with the jump rates (left), and the projected graph $\tilde{G}$ with the induced jump rates (right).}\label{piedi}
\end{figure}

  We can finally define the affinity $\cA (\cC)$ of a cycle $\cC$. To this aim recall that the CTRW is defined in terms of the dynamical characteristics $\psi_x$ and $p(x,y)$ and that we are considering also non  Markov CTRWs.
 Note that the jump probabilities $p(\cdot, \cdot)$ defined on $\cG$ can be projected on the graph $\tilde G$ without any ambiguity since we are assuming $n \geq 3$ (see 
 Fig. \ref{piedi} for an example).  
Finally, recall the definition of cycle affinity $\cA(\cC)$ (cf. \eqref{panda_grande}).
  
  \medskip

Let us now go back to the dynamics. Since  $X_0=\underline{v}^{(0)}$ we have $Y_0= v_*$ (recall that $Y_s= \pi(X_s) $). We now associate to each trajectory  $(X_s) _{s \in [0,t]}$ a cycle $\cC_t$ in $\tilde{G}$ as follows. Consider the projected path $(Y_s)_{s \in [0,t]}$. If $Y_t=Y_0= v_*$, then $\cC_t$ is given by the string of vertices visited by $(Y_s)_{s \in [0,t]}$,  taken in chronological order. If $Y_t \not = Y_0$, then we complete the above string by adding a path in $\tilde{G}$ from $Y_t $ to $v_*$ (this  additional  path depends only on $Y_t$: the same final point $Y_t$, the same additional path).  Finally, we take  the decomposition of the  random cycle $\cC_t$ in our fixed basis, i.e. 
 \begin{equation}\label{scandalo}
 \cC_t=  \sum_{i=1}^m a_i(t) \cC_i\,.\end{equation}
The fundamental link between the above construction and the original skeleton process is given by the following formula: \begin{equation}\label{sanpietro} | X_t^* -a_1(t) | \leq 1\,.
\end{equation} This is obtained observing that, since the graph $G$ is $(\underline{v}, \overline{v})$--minimal, it holds    $N_{v_*,z_1} (\cC_i)= \delta_{1,i}$ (and therefore $N_{v_*,z_1} (\cC_t)= a_1(t)$ by \eqref{scandalo}), and that   
 $N_{v_*,z_1}  (\cC_t)$ differs from $X_t^*$ by at most $1$.
 
 Having clarified the content of Theorem \ref{cicli}, we refer to   Appendix \ref{proof_teo_cicli} for its proof.

\section{Previous results on the asymptotic velocity and  large deviations}\label{panettone}
In this section we review some results of \cite{FS,FS1}. We point out that a key ingredient in their derivation has been the regenerative structure discussed in Section \ref{spa}. In Sections \ref{RW_Z_exp} and \ref{RW_Z_gamma} we will discuss  specific random walks for which  the LD rate functions entering in Theorem \ref{baxtalo} below  can be computed. On the other hand, Theorem \ref{giova} below will be very useful in the rest of the paper.

Recall that  $ X_0=\underline{v}^{(0)}$ and that 
(cf. Section \ref{spa})  $S_1$ denotes the first jump time for the skeleton process $X^*_t$, i.e.
	 \begin{equation}\label{pietro}
	 S_1 := \inf \left\{ t \geq 0 \,:\, X_t  \in \{ \underline{v}^{(-1)} , \underline{v}^{(1)} \}    \right \}\,.
  \end{equation}
 Recall also that we have assumed 
  that all the waiting times of $X_t$ have finite mean, i.e. $\psi_x$ has finite mean for all vertices $x$ in $G$. It is then simple to show that $\bbE (S_1) <\infty$. 
As derived in \cite{FS}, since   $\bbE (S_1) <\infty$, almost surely  the skeleton process  and therefore also the cell process admit  an asymptotic velocity:
\begin{equation}
\lim _{t \to \infty} \frac{N_t}{t}=
\lim _{t \to \infty} \frac{X^*_t}{t}=
 \frac{  \bbP (X^*_{S_1}=1)-\bbP (X^*_{S_1}=-1)}{\bbE(S_1)}=:v_{\lim}\,.\label{trenino}
\end{equation}
We refer the interested reader to \cite{FS} for what concerns the Gaussian fluctuations of $X^*$. In the rest of this section  we  concentrate on  large deviations.

\begin{TheoremA}[\cite{FS1}]\label{baxtalo}
 Call   $T_n$ the first time the skeleton process hits $n\in \bbZ$, i.e.
 \begin{equation}\label{chejo}
 T_n := \inf\left\{ t \geq 0 \,:\, X_t^* =n \right\} \in [0,+ \infty]\,.
 \end{equation}
 Then  the following holds:
 
 \begin{itemize}
 \item[(i)]
 As $n \to  \pm  \infty$ the random  variables  $T_{\pm n}/ n$ 
satisfy a LDP with speed $n$ and convex rate function
 \begin{equation}\label{pocoyo1}
 J_\pm (u):= \sup_{ \lambda \in \bbR} \left\{ \lambda u - \ln \varphi_\pm (\lambda) \right\}\,, \qquad u \in \bbR \,, \end{equation}
where  
\begin{equation} \label{pocoyo2}
 \varphi_\pm (\lambda ):= \bbE\left( e^{\lambda T_{\pm 1}} \mathds{1}(T_{\pm 1} <\infty )\right) \in (0,+\infty]\,, \qquad \l \in \bbR\,.
\end{equation}

\item[(ii)]
As $t \to +\infty$,  the random variables   $X^*_t /t$ and $N_t/t$ satisfy a LDP with speed $t$ and good\footnote{The rate function $I$ is good in the sense that $\{ \th : I(\th) \leq  a\}$ is compact, for any $a \in \bbR$}  and convex rate function $I$
given by 
\begin{equation}\label{rate_f}
I(\th)= 
\begin{cases}
\th J_+ (1 / \th) & \text{ if }\qquad  \th >0\,,\\
|\th |J_-(1 / |\th |) & \text{ if }\qquad  \th  <0\,,\\
\end{cases}
\end{equation}
and $I(0) = \lim _{\th \to 0} I(\th )$.
\end{itemize}
\end{TheoremA}
Roughly,  we have 
\[\bbP\left( \frac{T_{\pm n}}{n} \approx u \right)\sim  e^{- n J_\pm (u) } \,,\qquad \bbP\left(  \frac{N_t }{t} \approx \th \right) \sim e^{- t I(\th ) }\,,\qquad \bbP\left(  \frac{X^*_t }{t} \approx \th \right) \sim e^{- t I(\th ) }
\]
for $n,t$ large, respectively. 
 
It is useful for applications to reduce the computation of $\varphi_{\pm}$ to the one of  simpler functions. The following characterization of $\varphi_\pm$  is provided in \cite{FS1}, Proposition 4.3. Recall the definition of $S_1$ given in \eqref{pietro}, and let  $f_\pm: \bbR \to (0,+\infty]$ be defined by 
\begin{equation}\label{cioccolato}
   f_\pm (\lambda ) := \bbE \big( e^{\lambda S_1} \mathds{1}(X^*_{S_1} =\pm 1) \big) \in (0,+\infty] \,.
 \end{equation}
Then  the functions $\varphi_\pm(\l)$ in \eqref{pocoyo2} satisfy
\begin{equation} \label{cena}
\varphi_\pm (\l) = \frac{ 1 - \sqrt{ 1- 4 f_-(\l)f_+(\l) } }{ 2 f_\mp (\l) }
\end{equation}
for $\l \leq \l_c$, where $\l_c\in [0, +\infty )$ is the unique value of $\l$ such that $ f_-(\l)f_+(\l) = 1/4$, while $\varphi_\pm (\l) =+\infty$ for $\l > \l_c$.\\

In addition to Theorem \ref{lego_chima} the following characterization of the GC symmetry for the cell process 
is provided in \cite[Theorems 4,8]{FS1}:

\begin{TheoremA}[\cite{FS1}]\label{giova}
The following  facts are equivalent:
\begin{itemize}
\item[(i)] For some  $c \in \bbR$  the GC  symmetry
$I( \th)=I(-\th)-c \th$
holds for all $\th \in \bbR$;
\item[(ii)] The random variables $X_{S_1}$ and $S_1$ are independent. 
\item[(iii)]  The functions $\varphi_+(\l)$ and $\varphi_-(\l)$ are proportional, i.e.\ $\exists C>0$ such that $\varphi_+(\l)= C \varphi_-(\l)$ for all $\l \leq \l_c$. 
\end{itemize}
Moreover, when (i),(ii) hold it must be $c= \ln \frac{\bbP (X^*_{S_1}=1)}{\bbP (X^*_{S_1}=-1)}= \ln C$.
\end{TheoremA}

\begin{Remark} As a byproduct of Theorem \ref{giova} with Theorem  \ref{castelluccio} (or, equivalently, with \eqref{lorenza} in Theorem \ref{cicli}) the independence stated in Theorem \ref{giova}--(ii) holds for any $(\underline{v}, \overline{v})$--minimal fundamental graph.
\end{Remark}

\begin{Remark}\label{Arquata}
Note that, by \eqref{cena}, Item (iii) is equivalent to the proportionality of $f_+(\l)$ and $f_-(\l)$, which is often easier to check. Indeed, for $\l \leq \l_c$,  $\varphi_+(\l)= C \varphi_-(\l)$ if and only if $f_+(\l)= C f_-(\l)$.
\end{Remark}

\begin{Remark}\label{natale100} Consider a  CTRW  $X$   on $\bbZ$ with   $N$--periodic rates. Since the fundamental graph is given by a  finite linear graph and is therefore $(\underline{v}, \overline{v})$--minimal, we  know  that  the GC symmetry of Theorem \ref{giova}--(i) is satisfied (cf. Theorems \ref{castelluccio} and \ref{cicli}). As byproduct with Theorem \ref{giova} we 
get in particular that the time needed by $X$ to hit the set  $\{-N,N\}$ does not depend on which site is visited when first arriving in $\{-N,N\}$.  This property was already derived in \cite{DB} for CTRWs on $\bbZ$ with   $N$--periodic rates. \end{Remark} 

	\section{Example: Violation of GC symmetry with a non  $(\underline{v},\overline{v})$--minimal fundamental graph}\label{colomba}
	Considering a Markov CTRW, the violation of the GC symmetry for almost any choice of jump rates in the case of non 
	$(\underline{v},\overline{v})$--minimal fundamental graphs has a non trivial derivation, based on complex analysis \cite{FS1}. We discuss here an example, given by a parallel--chains model  \cite{K1}, confirming the thesis.
	
\begin{figure}[!ht]
    \begin{center}
     \centering
  \mbox{\hbox{
  \includegraphics[width=0.85\textwidth]{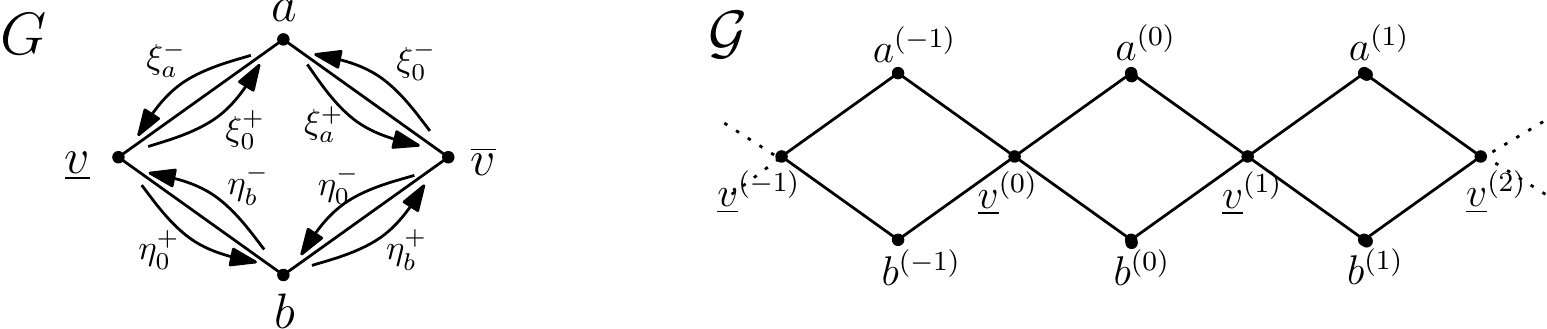}}}
         \end{center}
         \caption{The graph $G$ together with the rates (left) and the associated quasi--1d lattice $\cG$ (right).}   \label{koala}  
  \end{figure}

Let us consider the fundamental graph $G$ in Fig.\ \ref{koala} (left), in which to each pair of neighbouring vertices we have assigned a positive rate in $\{ \xi_0^\pm , \xi_a^\pm , \eta_0^\pm , \eta_b^\pm \}$. The associated quasi--1d lattice is represented in Fig.\ \ref{koala} (right).

Let $X_t$ denote the Markov CTRW on $\cG$ with periodic jump rates induced by  $G$. Finally, let $N_t$ and $X^*_t$ denote the cell process and skeleton process associated to $X_t$. 

By Theorem \ref{baxtalo}, as $t \to +\infty$,  the random variables $N_t/t$ and   $X^*_t /t$ satisfy a LDP with speed $t$ and rate function $I$, defined in \eqref{rate_f}. Since the fundamental graph $G$ is not $(\underline{v},\overline{v})$--minimal, we aim to show that $I$ satisfies the GC symmetry \eqref{gici} only for a set of transition rates $\{ \xi_0^\pm , \xi_a^\pm , \eta_0^\pm , \eta_b^\pm \}$ of zero Lebesgue measure in $[0,\infty )^{8}$. According to  Theorem \ref{giova}, to this end it suffices to show that $\varphi_+(\l)$ and $\varphi_-(\l)$ are not proportional for almost any choice of the jump rates. 

The computation of the ratio $\varphi_+(\l)/\varphi_-(\l)$ can  be reduced to a single cell analysis as follows.  Let $J_1$ be the first time that the process $X_t$, starting at $\underline{v}^{(0)}$, reaches $\{\underline{v}^{(-1)} , \underline{v}^{(0)} , \underline{v}^{(1)} \}$ after performing at least one jump, and set 
	\[ \tilde{f}_\pm (\l)   := \bbE ( e^{\l J_1} \mathds{1} (X^*_{J_1} = \pm 1  ) ) \,.\]
 One can  check that   $ 	\varphi_+(\l) = C\varphi_-(\l) $ if and only if $ \tilde{f}_+ (\l) =C \tilde{f}_- (\l) $ (see the beginning of Appendix \ref{castelluccio_proof} for more details). The advantage of introducing the auxiliary functions  $\tilde{f}_\pm(\l)$ is that in the present  parallel chain model they are simple to compute. 
  
Let $\t_1$, $\t_2$ denote  the first and second jump time of the process $X_t$ respectively, and observe that 
  	\begin{align*}
	& \{ X^*_{J_1}=1 \} = \{ X_{\t_1} = a^{(0)}, \, X_{\t_2} = \underline{v}^{(1)}  \} \cup \{ X_{\t_1} = b^{(0)}, \, X_{\t_2} = \underline{v}^{(1)} \} \, , \\
	& \{ X^*_{J_1}=-1 \} = \{ X_{\t_1} = a^{(-1)}, \, X_{\t_2} = \underline{v}^{(-1)}  \} \cup \{ X_{\t_1} = b^{(-1)}, \, X_{\t_2} = \underline{v}^{(-1)}  \} \, .
	\end{align*}
Hence, for $\l < \Psi_0, \Psi_a,\Psi_b$, it holds
	\[ \begin{split}
	\bbE ( e^{\l J_1} \mathds{1} (X^*_{ J_1} = 1 )) & = \sum _{z=  a^{(0)},b ^{(0)}}
	\bbE ( e^{\l (\t_1 + \t_2) } \mathds{1} (X_{\t_1} = z , \, X_{\t_2} = \underline{v}^{(1)} )) 
	\\ & 
	=\frac{ \xi_0^+ \xi_a^+}{(\Psi_0 - \l)(\Psi_a -\l)} + \frac{\eta_0^+ \eta_b^+}{(\Psi_0 - \l)(\Psi_b -\l)}  \,,
	\end{split} \]
where we have set $\Psi_0 := \xi_0^+ + \xi_0^- + \eta_0^+ + \eta_0^-$, $\Psi_a := \xi_a^+ + \xi_a^-$, $\Psi_b := \eta_b^+ + \eta_b^-$, and used that the waiting times at $\underline{v}^{(0)} , a^{(0)},b^{(0)}$ are exponentially distributed with inverse mean $\Psi_0, \Psi_a, \Psi_b$, respectively. If $\l$ is non smaller than $ \Psi_0, \Psi_a,\Psi_b$, then the expectation 
 $\bbE ( e^{\l J_1} \mathds{1} (X^*_{ J_1} = 1 )) $ diverges. 
 
  Repeating the same procedure for $\bbE ( e^{\l J_1} \mathds{1} (X^*_{J_1} = -1 )) $ we end up with 
	\[ 
	\tilde{f}_\pm(\l)  =
	\begin{cases}
	 \frac{ \xi_0^\pm \xi_a^\pm }{(\Psi_0 - \l)(\Psi_a -\l)} + \frac{\eta_0^\pm \eta_b^\pm}{(\Psi_0 - \l)(\Psi_b -\l)} & \text{ if } \l < \Psi_0, \Psi_a,\Psi_b \,,\\
	 +\infty & \text{ otherwise}   \,.
	 \end{cases}
	\]
We take the quotient to test proportionality. After a short calculation we find for  $\l < \Psi_0, \Psi_a,\Psi_b $:
	\[ \frac{ \tilde{f}_+(\l)}{\tilde{f}_-(\l)} = 
	\frac{ \xi_0^+ \xi_a^+ \Psi_b + \eta_0^+ \eta_b^+ \Psi_a - \l ( \xi_0^+ \xi_a^+ + \eta_0^+ \eta_b^+ )}
	{ \xi_0^- \xi_a^- \Psi_b + \eta_0^- \eta_b^- \Psi_a - \l ( \xi_0^- \xi_a^- + \eta_0^- \eta_b^- )} \, . \]
This is equal to a constant (independent of $\l$)  if and only if 
	\begin{equation} \label{determinant}
	 \det 
	\left( \begin{matrix}
	\xi_0^+ \xi_a^+ \Psi_b + \eta_0^+ \eta_b^+ \Psi_a & -(\xi_0^+ \xi_a^+ + \eta_0^+ \eta_b^+) \\
	\xi_0^- \xi_a^- \Psi_b + \eta_0^- \eta_b^- \Psi_a  & -  ( \xi_0^- \xi_a^- + \eta_0^- \eta_b^- )
	\end{matrix} \right)
	=0 
	\end{equation}
i.e. if and only if 
$ (\Psi_b - \Psi_a ) \big(  \xi_0^+ \xi_a^+  \eta_0^- \eta_b^-  -  \xi_0^- \xi_a^-  \eta_0^+ \eta_b^+ \big) = 0$.
Note that the second term vanishes if and only if $\frac{ \xi_0^- \xi_a^-}{\xi_0^+ \xi_a^+} = \frac{\eta_0^- \eta_b^-}{\eta_0^+ \eta_b^+ }$, and therefore \eqref{determinant} only holds on a set of jump rates of zero Lebesgue measure. We conclude that  for almost all choices of the jump rates  the functions $\tilde{f}_+$ and $\tilde{f}_-$ are not proportional, and therefore  the GC  symmetry \eqref{gici} is violated.


\section{Example: non  $(\underline{v} , \overline{v})$--minimal fundamental graph $G$ where the GC symmetry \eqref{gici} holds for any discrete time RW }\label{tagada}
We take the fundamental graph $G$ exactly as in Section \ref{colomba}. Moreover, 
we  take $\psi_x=\d_1$, hence the random walk jumps at each integer time.  Now  take  $\{ \xi_0^\pm , \xi_a^\pm , \eta_0^\pm , \eta_b^\pm \}$  to be  jump probabilities: $\xi_0^++\xi_0^-+\eta_0^++\eta_0^-=1$, $\xi_a^-+\xi_a^+=1$ and $\eta_b^-+\eta_b^+=1$.
As discussed in the previous section, 	to prove that the GC symmetry \eqref{gici} is  satisfied  for any choice of the jump probabilities, 
it is enough to show that $\tilde f_+(\l)=C \tilde f_-(\l)	$. In this case (see Theorem \ref{giova} and Remark \ref{Arquata}), \eqref{gici} is satisfied with $c=\ln C$. Since trivially 
\[ \tilde f _+(\l) = e^{2\l} ( \xi_0^+\xi_a^++\eta_0^+\eta_b^+)\,,\qquad \tilde f _-(\l) = e^{2\l}( \xi_0^-\xi_a^-+\eta_0^-\eta_b^-)
\,,
\]
we conclude that \eqref{gici} is always satisfied with $c=\ln\frac{ \xi_0^+\xi_a^++\eta_0^+\eta_b^+}{\xi_0^-\xi_a^-+\eta_0^-\eta_b^-}  $.

\section{Example: homogeneous CTRW on $\bbZ$ with exponentially distributed waiting times }\label{RW_Z_exp}
\begin{figure}[!ht]
    \begin{center}
     \centering
  \mbox{\hbox{
  \includegraphics[width=0.7\textwidth]{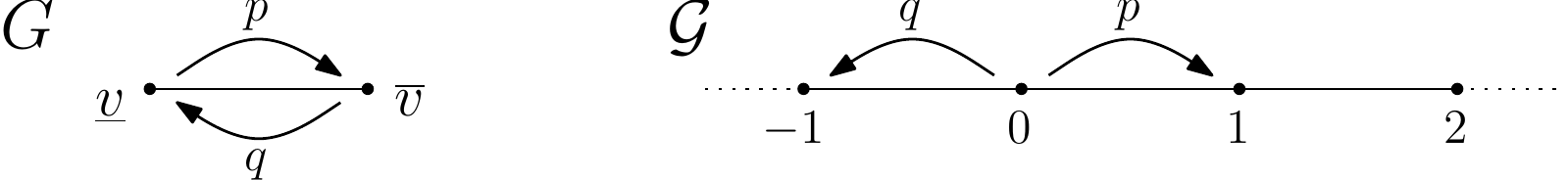}}}
         \end{center}
         \caption{The fundamental graph $G$ (left) and the associated quasi--1d lattice $\cG$ (right).}\label{simple_pic}
  \end{figure}

In this section we consider the simplest possible $(\underline{v} , \overline{v})$--minimal fundamental graph $G$, given by only two vertices $\underline{v} , \overline{v}$, connected by an edge. We assign rate $p\in (0,1)$ to the oriented edge $(\underline{v} , \overline{v})$ and rate $q=1-p$ to the reverse edge $(\overline{v} , \underline{v})$, as represented in Fig.\ \ref{simple_pic}. 

Let $X_t$ be a Markov CTRW (with rates $p,q$) on the associated quasi--1d lattice $\cG$, whose vertex set we identify with $\bbZ$. Then $X_t$ waits at each location an exponentially distributed time of mean $1$, and then jumps either to the right or to the left with probability $p$ and $q$ respectively. Note that in this case  the skeleton process $X^*_t$ coincides with the random walk $X_t$. 

Our aim here it to implement Theorem \ref{baxtalo} and derive  explicit expressions  for  the LD rate functions $J_\pm$ of the hitting times and for the LD rate function $I$ of  the cell process.

As usual we take  $X_0=0$. Then, as it is well known and can be also recovered from the formulas in \cite{FS}, the asymptotic velocity and diffusion coefficients of $X_t$ are respectively given by  $v_{\lim} = p-q $ and $\s^2 =1$. Moreover, by Theorem \ref{baxtalo}, as $t\to\infty$ the process $X_t/t$ satisfies a LDP with speed $t$ and good and convex rate function $I$. In this section we show how $I$ can be computed using our approach. 

Recall that $S_1$ denotes the first jump time for the skeleton process, which in this case coincides with the first jump time of $X_t$. Then the functions $f_\pm$ defined in \eqref{cioccolato} are given by	
	\begin{align*}
	& f_+(\l ) = \bbE \big( e^{\lambda S_1} \mathds{1}(X_{S_1} = 1) \big)
	=  	\begin{cases}
	\frac{p}{1-\l } \, , \quad \l < 1 \\
	+ \infty \, , \mbox{ otherwise.} 
	\end{cases} \\
	& f_-(\l ) = \bbE \big( e^{\lambda S_1} \mathds{1}(X_{S_1} = -1) \big)
	 = 
	\begin{cases}
	\frac{q}{1-\l } \, , \quad \l < 1 \\
	+ \infty \, , \mbox{ otherwise.} 
	\end{cases}
	\end{align*}
Hence,  $f_+(\l ) f_-(\l) = 1/4 \Leftrightarrow \l = 1 - 2\sqrt{pq} =: \l_c$. Moreover, by \eqref{cena}, for $\l\leq \l_c$
we have
	\[ \f_+ (\l ) = \frac{1-\l - \sqrt{ (1-\l)^2 - 4pq}}{2q} \, ,  \qquad
	  \f_- (\l ) = \frac{1-\l - \sqrt{ (1-\l)^2 - 4pq}}{2p} = \frac{q}{p} \f_+(\l) \, , \]
	  otherwise $\varphi_\pm (\l)=\infty$.
It follows that $\ln \f_+ (\l) = \ln \f_- (\l) - (\ln q/p )$ which, together with  \eqref{pocoyo1}, gives  $J_+ (\th ) = J_-(\th) + (\ln q/p ) $ for all $\th \in \bbR$. 
This implies that the rate function $I(\th)$ satisfies the GC symmetry $I(\th ) - I(-\th) = c \th$, the constant $c$ being indeed $\ln{q/p}$ (which we already knew from Theorem \ref{lego_chima}).

In order to compute $I$, we note that  
	\[ \frac{\rm d}{{\rm d} \l} \ln \f_\pm (\l) = \frac{1}{\sqrt{ ( 1-\l)^2 - 4pq}}\,, \qquad \mbox{ for }\l < \l_c \,.\]
Solving $\frac{\rm d}{{\rm d} \l} \ln \f_\pm (\l) = \th$ for $\th > 0$, we find $\tilde{\l}_\pm (\th) = 1 - \frac{1}{\th} \sqrt{ 1 + 4pq \th^2 }$ and
	\[ \begin{split}
	J_+ (\th ) & = \th \tilde{\l}_+ (\th) - \ln \f_+ ( \tilde{\l}_+ (\th) ) 
	\\ & = \th - \sqrt{ 1+4pq \th^2} + \ln (2\th q) - \ln ( \sqrt{ 1 + 4pq \th^2 } -1) 
	\end{split} \]
for $\th >0$, $J_+(\th) = +\infty$ otherwise. This also gives us an explicit expression for $J_- (\th ) = J_+(\th) - (\ln q/p ) $. 
Note that $\lim_{\th \searrow 0 }J_\pm (\th) = +\infty$ and $J_\pm $ has a critical point (i.e.\ with vanishing derivative) at $\th = \frac{1}{\sqrt{1-4pq}}$ which is infinite if and only if  $p=q=1/2$. When finite, $ \frac{1}{\sqrt{1-4pq}}$ is a point of minimum for $J_\pm$. 
Recalling the definition of the rate function $I$ given in \eqref{rate_f}, then, we conclude that $I$ is finite for all $\th \in \bbR$ and 
	\begin{equation}\label{zucchero} I(\th) = 1 - \sqrt{ \th^2 + 4pq} - \th \ln ( \sqrt{\th^2 + 4pq} -\th ) + \th \ln (2q) \, . \end{equation}
Finally, we point out that, since 
\[  - \th \ln ( \sqrt{\th^2 + 4pq} -\th ) + \th \ln (2q)= \th \ln ( \sqrt{ \th^2 + 4pq} +\th )- \th \ln 2p \,,
 \]
 formula \eqref{zucchero}  matches  equation (10.6)  in \cite{BFG2}. 
 
\begin{figure}[!ht]
\begin{minipage}[b]{0.48\linewidth}
\centering
\includegraphics[width=\textwidth]{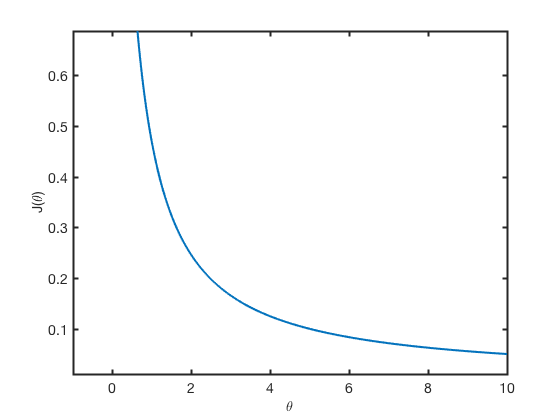}
\includegraphics[width=\textwidth]{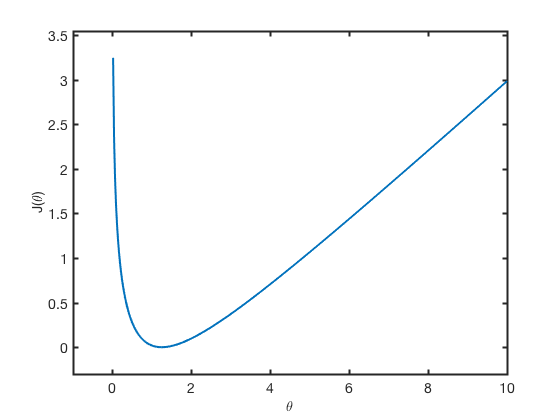}
\end{minipage}
\begin{minipage}[b]{0.48\linewidth}
\centering
\includegraphics[width=\textwidth]{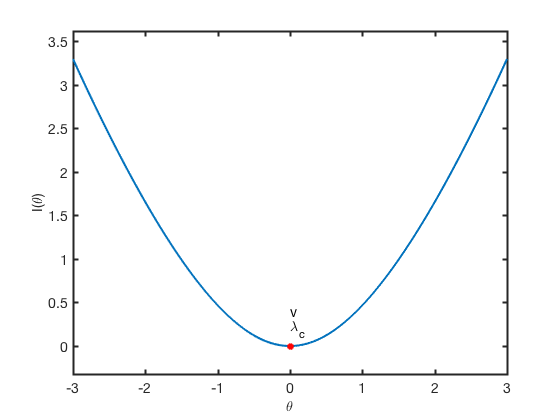}
\includegraphics[width=\textwidth]{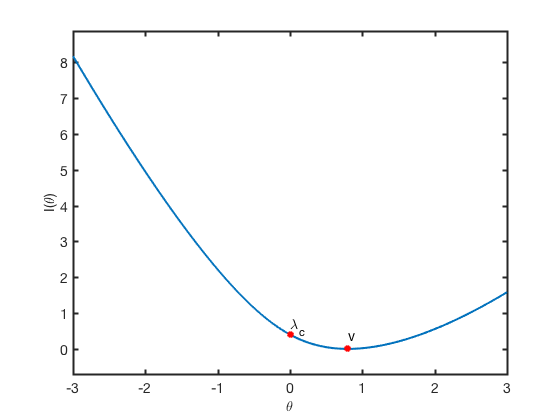}
\end{minipage}
\caption{Plots of the rate functions $J_+$ (left) and $I$ (right). The top line concerns the symmetric case $p=q=0.5$, for which $v=\l_c=0$. The bottom line concerns the asymmetric case $p=0.9$, $q=0.1$, for which $v=0.8$ and $\l_c = 0.4$. }
\end{figure}


 \section{ Example: homogeneous CTRW on $\bbZ$ with Gamma--distributed waiting  times }\label{RW_Z_gamma}

In this section we again consider the very simple fundamental graph $G$ in Fig.\ \ref{simple_pic}, together with the associated quasi--1d lattice $\cG$. This time, on the other hand, we assume that the CTRW $X_t$ on $\cG$  waits at each location $x\in \bbZ$ a Gamma--distributed random time (non exponential) and then jumps to either $x+1$ or $x-1$ with probability $p$ and $q=1-p$  respectively. 

Note that again $X_t$ coincides with the associated skeleton process $X_t^*$. Moreover, since the waiting times are not exponentially distributed, the process is   not Markovian. Our aim here it to implement Theorem \ref{baxtalo} and derive  explicit expressions  for  the LD rate functions $J_\pm$ of the hitting times and for the LD rate function $I$ of  the cell process. This can indeed be achieved when the waiting times have distribution $\textrm{Gamma}(2,\g)$  for some $\g>0$.

Assume $X_0=0$ as usual, and note  that $S_1$ introduced in \eqref{pietro}  equals  the first jump time of the process $X_t$. We first assume that  $S_1\sim \textrm{Gamma}(\n,\g)$ for some parameters $\nu , \gamma >0$, which means that the probability density function of $S_1$ is of the form $f(t) \propto t^{\n-1} e^{-\g t}$ for $t>0$. Moreover,  it holds $\bbE (e^{\l S_1}) = \big( 1 - \frac{\l}{\g} \big)^{-\n}$ for $\l < \g$, $\bbE (e^{\l S_1}) = \infty$ otherwise. Note that if $\n =1$ we are back to exponential holding times of parameter $\g$. 


 Recall that, according to \eqref{rate_f} in Theorem \ref{baxtalo}, $I(\th )$ for $\th >0$ can be deduced from the hitting times rate function $J_+$, defined in \eqref{pocoyo1} as 
	\begin{equation} \label{J2}
	 J_+(u) := \sup_{\l \in \bbR} \{ \l u - \ln \f_+(\l) \} \, . 
	 \end{equation}
The other branch of $I$ is then easily obtained by mean of the GC symmetry (see Theorem \ref{castelluccio} or  \eqref{lorenza} in Theorem \ref{cicli}). 

In order to compute $J_+$ we observe that, using the independence of $S_1$ and $X_{S_1}$ (which is a byproduct of e.g. Theorem \ref{castelluccio} with Theorem \ref{giova}), for  $\l < \g$ we have
	\begin{align*}
	&  f_+(\l) = \bbE(e^{\l S_1}) \bbP(X_{S_1}=1) = p \Big( 1 - \frac{\l}{\g} \Big)^{-\n}\,,\\& 
	f_-(\l) = \bbE(e^{\l S_1}) \bbP(X_{S_1}=-1) = q \Big( 1 - \frac{\l}{\g} \Big)^{-\n} \, . \end{align*}
	Solving $f_+(\l)f_-(\l) = 1/4$, then, we find
$ \l_c = \g \big( 1 - (4pq)^{1/2\n }\big)  $ 
and, by \eqref{cena},
	\[ \f_+(\l) = \frac{1-\sqrt{1-4f_+(\l)f_-(\l)}}{2f_-(\l)} 
	= \frac{ \big( 1 - \frac{\l}{\g} \big)^\n - \sqrt{\big( 1 - \frac{\l}{\g} \big)^{2\n} - 4pq}}{2q}  \]
for $\l \leq \l_c$, $\f_+(\l) = \infty$ otherwise.

Let us now compute the supremum in \eqref{J2}. According to \eqref{rate_f} we are only  interested in $J_+(u)$ for $u>0$, that we assume throughout.
Observe that the supremum can be restricted to $\l \leq \l_c$, since $\ln \f_+ = \infty$ otherwise. 
Moreover, for $\l < \l_c$ we can differentiate the argument, to find that the supremum is attained at $\l (u)$ solution of $\f_+'(\l) = u \f_+(\l)$.  Since
	\[ \f_+'(\l) = \frac{1}{2q} \bigg( -\frac{\n}{\g} \bigg) \bigg( 1-\frac{\l}{\g}\bigg)^{\n -1} 
	\Bigg[ 1 - \frac{  \big( 1-\frac{\l}{\g}\big)^\n}{ \sqrt{ \big( 1-\frac{\l}{\g}\big)^{2\n} - 4pq}} \Bigg] \, , \]
the equation $\f_+'(\l) = u \f_+(\l)$ reads
	\[ u \sqrt{ \bigg( 1-\frac{\l}{\g}\bigg)^{2\n} - 4pq} = \frac{\n}{\g} \bigg( 1 - \frac{\l}{\g} \bigg)^{\n-1} .\]
This can be explicitly solved for $\nu=2$, to get 
	\begin{equation}\label{video}  \l (u)= \g \Bigg[ 1 - \frac{1}{u} \sqrt{ \frac{2}{\g^2} + \sqrt{ \frac{4}{\g^4} + 4pqu^4}} \Bigg] \, . \end{equation}
From now on we assume $\nu=2$. Plugging \eqref{video} back into \eqref{J2}, we find
	\[ \begin{split} 
	J_+(u) & = u \l (u) - \ln \f_+(\l(u))   = 
	\g u \Bigg[ 1 - \frac{1}{u} \sqrt{ \frac{2}{\g^2} + \sqrt{ \frac{4}{\g^4} + 4pqu^4}} \Bigg] + \\ &
	- \ln \Bigg[ \frac{1}{u^2} \bigg( \frac{2}{\g^2} + \sqrt{ \frac{4}{\g^4} + 4pqu^4} \bigg) 
	- \frac{2}{\g u^2} \sqrt{ \frac{2}{\g^2} + \sqrt{ \frac{4}{\g^4} + 4pqu^4}}  \Bigg] 
	+ \ln(2q) \, . 
	\end{split}\]
Using now that $I(u) = uJ_+(1/u) $ for $u>0$ by \eqref{rate_f}, we conclude that
	\[ \begin{split} I(u) & =
	 \g  \Bigg[ 1 - u \sqrt{ \frac{2}{\g^2} + \sqrt{ \frac{4}{\g^4} + \frac{4pq}{u^4}}} \Bigg] + \\ &
	- u \ln \Bigg[ u^2 \bigg( \frac{2}{\g^2} + \sqrt{ \frac{4}{\g^4} + \frac{4pq}{u^4}} \bigg) 
	- \frac{2 u^2}{\g} \sqrt{ \frac{2}{\g^2} + \sqrt{ \frac{4}{\g^4} + \frac{4pq}{u^4}}}  \Bigg] 
	+ u \ln(2q) \, 
	\end{split} \]
for $u>0$. Using the GC symmetry \eqref{lorenza}, which reads 
	\[ I(-u) = I(u) - u \ln (q/p) \, ,  \]
we also obtain $I(u)$ for $u<0$. Finally, $I(0) = \l_c = \g \big( 1 - \sqrt[4]{4pq} \big) $.

\begin{figure}[!ht]
    \begin{center}
     \centering
  \mbox{\hbox{
  \includegraphics[width=0.6\textwidth]{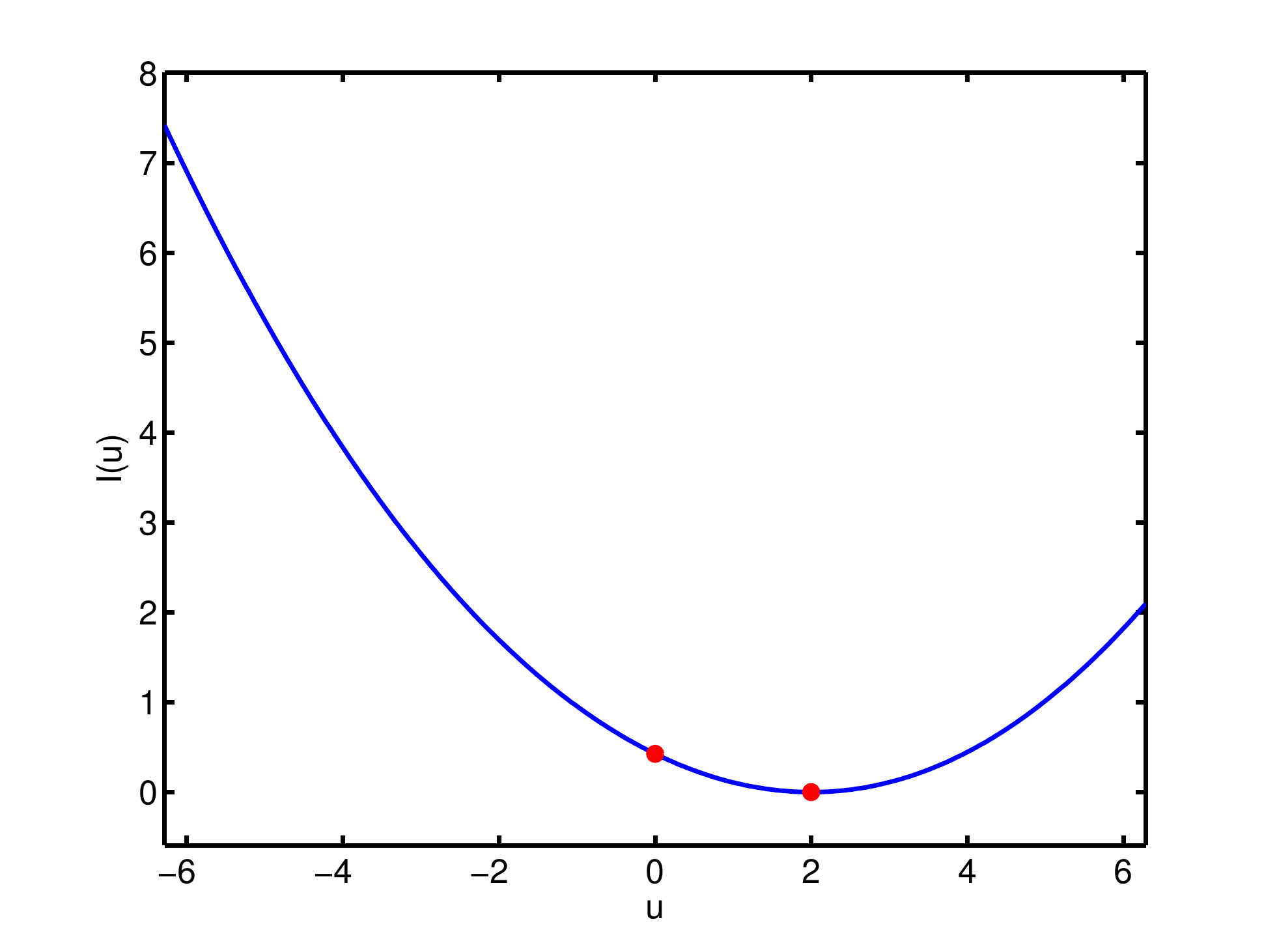}}}
         \end{center}
         \caption{The rate function $I(u)$ for $p=0.7$, $q=0.3$, $\n =2$, $\g = 10$. The red dots correspond to the points $(0,\l_c )$ and $(v_{\lim},0)$, where $v_{\lim} = \frac{p-q}{\bbE(S_1)} = \frac{(p-q)\g}{2} $ is the asymptotic velocity of the process $X_t$.}
  \end{figure}
   
   \appendix 
   

\section{Derivation of Criterion \ref{criceto} implying Theorem \ref{castelluccio}}\label{castelluccio_proof}
Recall that  $X_0= 
\underline{v}^{(0)}$. We define $J_1$ as the first time that, after at least one jump, the random walk visits again a state of the form $\underline{v}^{(k)}$, i.e.
\[J_1:= \inf\bigl\{ t >0\,:\, X_t \in \{\underline{v} ^{(-1)}, \underline{v}^{(0)} \underline{v}^{(1)}\} \text{ and }   \; \exists s \in (0,t) \text{ with } X_s \not = X_0 \bigr\}\,.\]
Moreover, we  set
	\begin{equation}\label{caffe}
	\tilde{f}_\pm (\l)   := \bbE ( e^{\l J_1} \mathds{1} (X_{J_1} = \underline{v}^{(\pm 1)}  )\, , \qquad 
	\tilde{f}_0 (\l)   := \bbE ( e^{\l J_1} \mathds{1} (X_{J_1} = \underline{v}^{(0)}) ) \, . 
		\end{equation}
			Due to   \cite[Lemma 9.1]{FS}  the following holds (recall \eqref{cioccolato}):
if $ \tilde f_0(\l) < 1$, then 
\[	f_+(\l) =
	\frac{ \tilde{f}_+(\l)}{1- \tilde{f}_0 (\l)} \,, \qquad 
	f_-(\l) = \frac{  \tilde{f}_-(\l)}{1-\tilde{f}_0 (\l)} \,.\]
	Moreover, if 
	$ \tilde f_0(\l) \geq1$,  then  $f_+(\l)= f_-(\l)= +\infty$.
	
	As  a byproduct with Theorem \ref{giova} and Remark \ref{Arquata} we conclude that the GC symmetry \eqref{verabila} is satisfied   for some constant $\D$ if and only if  $f_+(\l)= e^{\D} f_-(\l) $ for all $\l \leq \l_c $, and therefore if and only if  $\tilde{f}_+(\l)= e^{\D}\tilde{f}_-(\l)$ for all $\l \leq \l_c $.

 Given an integer $m \geq 1$, let $\cA_m$ be the family of strings $(x_0,x_1, \dots, x_m)$ such that
 $x_0= \underline{v}$, $x_m = \overline{v}$, $(x_i,x_{i+1}) $  is an edge of the fundamental graph $G$ for all $i:0\leq i <m$ and $ x_i \in V \setminus \{\underline{v}, \overline{v} \} $ for all $0<i < m$. We call $\cA_m^*$ the family of 
 sequences  satisfying the same properties as above  when exchanging the role of  $\underline{v}$ and $\overline{v}$. 

 Recall that $\psi_x $ is the law of the waiting time at $x$. We set
 \[ \phi_x(\l) := \int e^{\l t} \psi_x (dt )\]
 for its Laplace transform. 
 Then we can write
\begin{equation}\label{giaco0}
\tilde{f}_+(\l) = \sum_{m=1}^\infty \sum _{(x_0, x_1, \dots, x_m) \in \cA _m}  \prod _{i=0}^{m-1} p(x_i,x_{i+1}) \prod _{i=0}^{m-1} \phi_{x_i} (\l)
\,.
\end{equation}
 A similar expression holds for $\tilde{f}_-(\l)$, with $\cA_m$ replaced by $\cA_m^*$.

Given $\g= (x_0,x_1, \dots, x_m)$  we set 
$$ R_\g :=  \prod _{i=0}^{m-1} p(x_i,x_{i+1})   \,, \qquad S_\g(\l)= \prod _{i=1}^{m-1} \phi_{x_i}(\l) \,.
$$
Note that the product in $S_\g(\l)$ starts from $i=1$, hence all sites $x_i$ appearing in the product belong to $V \setminus \{ \underline{v}, \overline{v} \}$. By mean of this notation we can write 
\begin{equation}\label{modena}
\tilde{f}_+(\l) = \phi_{\underline{v}}(\l) \sum _{m=1}^\infty \sum _{\g \in \cA_m } R_\g  S_\g(\l) \,, \qquad \tilde{f}_-(\l) =   \phi_{\underline{v}}(\l)  \sum _{m=1}^\infty \sum _{\g \in \cA_m^* } R_\g  S_\g(\l)\,.
\end{equation}
Due to the previous observations,   the GC symmetry  \eqref{verabila} is satisfied   for some constant $\D$ if and only
\begin{equation}\label{nicholls}
\sum _{m=1}^\infty \sum _{\g \in \cA_m } R_\g  S_\g(\l)=  e^\D \sum _{m=1}^\infty \sum _{\g \in \cA_m^* } R_\g  S_\g(\l)\,,  \qquad \l \leq \l_c \,.\end{equation}

\begin{figure}[!ht]
    \begin{center}\label{rumiz}
     \centering
       \mbox{\hbox{
  \boxed{
  \includegraphics[width=0.45\textwidth]{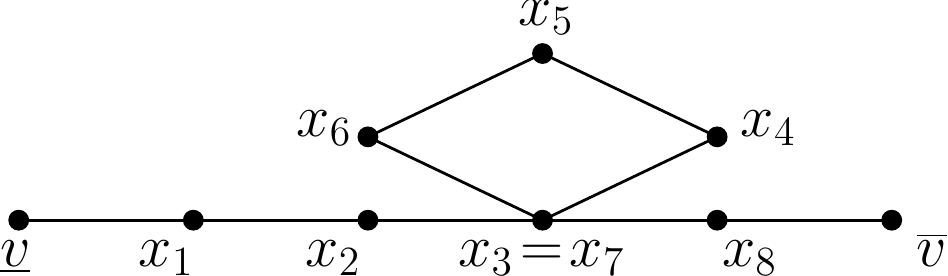}}}}
            \end{center}
            \caption{Example of a path $\g \in \cA_9$. $\bar \g =( \underline{v},
         x_1, x_2,x_3,x_8, \overline{v})$,   $R^{\rm loop}_{\gamma} = p(x_3,x_4) p (x_4,x_5) p (x_5,x_6) p(x_6,x_7)$, $\g^\dag=( \overline{v}, x_8,x_7, x_4,x_5,x_6,x_3,x_2,x_1, \underline{v})$.}
       \end{figure}

Given $\g= (x_0,x_1, \dots, x_m) \in \cA_m $  we write $\bar \g $ for the \emph{ loop--erased} version of $\g$ (see Fig. \ref{rumiz} taken from \cite{FS}). We recall that $\bar \g$ is obtained by   erasing all the loops of $\g$  in chronological order. More precisely, consider the following algorithm. Set $i_0 := 0$ and, once defined $i_0, i_1, \dots,i_k$, set $ r:= k$ if $i_k =m $, otherwise (if $i_k <m$) set 
\[  i_{k+1}:= \max\{ j:  i_k \leq  j < m  \text{ and } x_j= x_{i_k}\} +1\] (recall that $\g \in \cA_m$ visits $\overline{v}$ only as last point $x_m = \overline{v}$).  Then  the loop--erased version of $\g$ is given by  $\bar \g =( x_{i_0}, x_{i_1}, \dots, x_{i_r})$. Since $\g \in \cA_m$ it must be $\bar \g \in \cA_r$.  Note that
\begin{equation}\label{mauritania}
R_\g = R_{\bar \g}  R^{\rm loop}_{\gamma} \,, \quad
R^{\rm loop} _\g:=  \prod _{k=0}^{r-1}  \prod _{ i= i_k}^{i_{k+1}-2} p(x_i, x_{i+1})\,,
\end{equation}
with the convention that $\prod _{ i= i_k}^{i_{k+1}-2} p(x_i, x_{i+1}) =1$ if $i_{k+1} = i_k +1$
($R^{\rm loop}_\g $ is the contribution to $R_\g$ given by factors associated to  the edges inside the loops, see Fig. \ref{rumiz}).

We write $\g ^\dag$ for the path  in $\cA_m^*$ going from $\overline{v} $ to $\underline{v}$ and obtained from $\g$ by reversing the order out of the loops  and keeping the same order in the loops (see Fig. \ref{rumiz}).  
More precisely, with the notation introduced above, it holds
$$  \g^\dag =\bigl( x_{i_r},  \star \star \star   , x_{i_{r-1} },\star  \star \star, x_{i_{r-2} }, \star \star \star, x_{i_1}, \star\star  \star, x_{i_0} \bigr)\,.
$$
where the pieces  marked  by $\star \star \star$ are determined as follows.  Take $k: 1\leq k \leq  r$.
If $i_k= i_{k-1}+1$, 
then   the piece ``$ x_{i_k},  \star \star \star   , x_{i_{k-1} }$'' is indeed simply ``$x_{i_k}, x_{i_{k-1} }$''. If  $ i_k >
i_{k-1}+1$,  then  the piece ``$ x_{i_k},  \star \star \star   , x_{i_{k-1} }$'' is given by 
``$ x_{i_k},  x_{i_k -1} = x_{i_{k-1}} ,  x_{i_{k-1} +1},  x_{i_{k-1} +2} , \dots,  x_{i_k -1}  = x_{i_{k-1} }$''.


We note that the transformation $\g\mapsto \g^\dag$ is an involute bijection from $\cA_m$ to $\cA_m^*$ and that 
\begin{equation}\label{cud}
\overline{ ( \g ^\dag)}= (\bar \g )^\dag \,, \qquad 
R_\g^{\rm loop}= R_{\g^\dag} ^{\rm loop}\,, \qquad 
 S_\g (\l)= S_{\g^\dag}(\l)\,.
 \end{equation}
 
 The above identities, together with \eqref{nicholls} and \eqref{mauritania}, imply the following criterion:
 
 \begin{CriterionA}\label{criceto}
  The GC symmetry  \eqref{verabila} is satisfied   for some constant $\D$ if and only if  
\begin{equation}\label{berto}
\sum _{m=1}^\infty \sum _{\g \in \cA_m } R_\g^{\rm loop}\bigl( R_{\bar \g}- e^\D R_{ \bar \g^\dag} \bigr)S_\g(\l) =0
\end{equation} 
 for $\l \leq \l_c$.
In particular,   \eqref{verabila} is satisfied   for some constant $\D$ if it holds $ R_{\g}= e^\D R_{ \g^\dag}$ for any loop--free path $\g$ from $\underline{v}$ to $ \overline{v}$. 
 \end{CriterionA}
 The above criterion is similar to Criterion 1 in \cite[Sec. 9]{FS1}  and the derivation of  Criterion 1 is indeed inspired by \cite[Sec. 9]{FS1}.

  If $G$ is \emph{$(\underline{v}, \overline{v})$}--minimal, then trivially the above criterion implies that \eqref{verabila} is satisfied with $\D$ given by \eqref{alexey_biz}.

   \section{Derivation of Theorem \ref{cicli}} \label{proof_teo_cicli}
Recall that $\tilde{G}$ denotes the projected graph $\pi (G)$ as in Fig.\ \ref{piedi}, and $Y_t = \pi (X_t)$ is the projection of $X_t$ onto $\tilde{G}$. 
%

The first part of the theorem up to \eqref{crimea} is known (cf. \cite[Section 8]{F} and the theorem on \cite[page 7]{AG3}).
Equation  \eqref{mosca} is a byproduct of \eqref{crimea} and \eqref{ucraina}. Finally, since $I(\th)= \inf _{\th_2, \dots, \th_m} \cI( \th, \th_2, \dots, \th _m)$ by the contraction principle, the symmetry \eqref{lorenza} follows trivially from \eqref{mosca}.

%

 It therefore remains  to prove \eqref{ucraina}.
To this aim
recall that $( z_0 , z_1 , \ldots , z_n)$ denotes the unique self--avoiding path from $z_0 = \underline{v} $ to $ z_n =\overline{v}$ in $G$. 
A path $(x_s)_{s \in [0,t]}$ on $\tilde G$, starting at $v_*$, makes excursions outside the set $\{v_*, z_1,\dots, z_{n-1}\}$.  Since $G$ is  $(\underline{v}, \overline{v})$--minimal, the graph   $\tilde G$ is given by the cycle $\cC_1$ to which subgraphs $G_1, G_2, \dots, G_k$ are attached in such a way that $\cC_1$ and $G_i$ share a unique vertex (see Figure \ref{fig:incollo}).  We consider the transformed path $\G ( ( x_s)_{s \in [0,t]})$ where each excursion is replaced by a time inversion as follows. Suppose that at time $s_0$ the path 
$(x_s)_{s \in [0,t]}$ enters some subgraph  $G_i$ and it exits at some time $t_0>s_0$.
Then we replace the excursion  $(x_s)_{ s_0 \leq s \leq t_0}$ with the excursion $(x_{s_0+ (t_0-s) } )_{ s_0 \leq s \leq t_0}$, apart from some modifications at the jump times in order to have a c\`adl\`ag path at the end. To give a precise definition, call 
 $v$ the unique vertex in common between $G_i$ and $\tilde G$. Then  $x_{s_0}=x_{t_0}=v$. Let us suppose that during the excursion the path visits (in chronological order) the vertices  $y_0, y_1, \dots,y_{r-1},  y_r$ where $y_0= y_r =v$ and $y_1, \dots,y_{r-1} \in G_i \setminus \{v\}$, and that it remains at site $y_m $ a time $\t_m$ for any $m=0,1, \dots, r$. Note that it must be $ t_0= s_0+\t_0+\t_1+ \cdots + \t_{r}$. Then we replace the excursion $(x_s)_{ s_0 \leq s \leq t_0}$ by the path that visits  (in chronological order) the vertices   $y_r, y_{r-1}, \dots,y_{1},  y_0$  with consecutive waiting times given by $\t_r, \t_{r-1}, \dots,  \t_1, \t_0$ 
(i.e.\ at time $s_0$ the new path  is at $v=y_r$ where it remains for a  time 
$\t_r$, then the new path jumps to $y_{r-1}$ where it  remains for a  time $\t_{r-1}$ and so on 
until  jumping to  $y_0=v$ where it remains for a time $\t_0$ and then finally leaves the subgraph $G_i$ at time $s_0+\t_r+ \t_{r-1}+ \dots+ \t_2+ \t_1+ \t_0=t_0$).

  Since $\G$ is an involution, 
 it is simple to compute the Radon--Nykodim derivative $dP/dQ$. We claim that 
\begin{equation}\label{scappo!}
 \frac{dP}{dQ}  \left( (x_s)_{s \in [0,t]}\right) = \frac{  dP \left( (x_s)_{s \in [0,t]}\right) }{ dP \left( \G \left( (x_s)_{s \in [0,t]}\right)\right) }=e^{ \sum _{i=2}^m a_i(t) \cA( \cC_i)+O(1)} 
 \end{equation}
where $a_i(t)$ refers to the cycle $\cC_t$ associated to the path  $(x_s)_{s \in [0,t]}$.

To check the above formula suppose for simplicity that $x_0=x_t= v_*$ (hence $(x_s)_{s \in [0,t]}= \cC_t$). 
If  $(x_s)_{s \in [0,t]}$ visits (in chronological order) the sites $w_0,w_1, \dots, w_k$ with holding times (up to time $t$) given by $\t_0, \t_1, \dots, \t_k$ we have \[ dP \left( (x_s)_{s \in [0,t]}\right) = A \left( (x_s)_{s \in [0,t]}\right) B\left( (x_s)_{s \in [0,t]}\right)\]
where 
\[ A\left( (x_s)_{s \in [0,t]}\right)= \prod _{i=0}^{k-1} p(w_i, w_{i+1} )  \,,\qquad B\left( (x_s)_{s \in [0,t]}\right)= \left[\prod _{i=0}^{k-1} \psi_{w_i} (d \t_i ) \right]\psi _{w_k} ([\t_k, \infty))\,. \]
Note that $ B\left( (x_s)_{s \in [0,t]}\right)=  B\left( \Gamma\left( (x_s)_{s \in [0,t]}\right)\right)$. On the other hand,   we have 
\[ 
 \frac{  A\left( (x_s)_{s \in [0,t]}\right)}{  A \left( \Gamma\left( (x_s)_{s \in [0,t]}\right)\right)} = \prod _{ \substack{ i: 0\leq i <k \\ (w_i , w_{i+1}) \not\subset  \cC_1}  }\frac{ p(w_i, w_{i+1} )}{ p ( w_{i+1}, w_i)}\,.
\]
where the product in the r.h.s. is among the edges $ (w_i , w_{i+1})$ which are not edges of $\cC_1$, neither when  reversing orientation.
 To conclude, it remains to check that 
 \begin{equation}\label{calvez}\sum  _{ \substack{ i: 0\leq i <k \\ (w_i , w_{i+1}) \not\subset  \cC_1}  } \ln \frac{ p(w_i, w_{i+1} )}{ p ( w_{i+1}, w_i)}= \sum _{i=2}^m a_i(t) \cA( \cC_i)\,.
 \end{equation}
To this aim, for each unoriented edge not in $\cC_1$,  fix a canonical orientation and call $\cE$ the family of such canonically oriented edges. Then we have 
 \[   \sum  _{ \substack{ i: 0\leq i <k \\ (w_i , w_{i+1}) \not\subset  \cC_1}  } \ln \frac{ p(w_i, w_{i+1} )}{ p ( w_{i+1}, w_i)}
 =
\sum _{(x,y) \in \cE} N_{x,y} (\cC_t)\ln \frac{ p(x, y )}{ p (y,x)} \,.\] On the other hand, since $\cC_t = \sum _{i=1}^m a_i(t) \cC_i$ and since $N_{x,y}(\cC_1)=0$ for any $(x,y) \in \cE$, we have
\[   \sum _{(x,y) \in \cE} N_{x,y} (\cC_t)\ln \frac{ p(x, y )}{ p (y,x)}= 
\sum _{i=2}^m a_i(t)  \sum _{(x,y) \in \cE}  N_{x,y} (\cC_i)\ln \frac{ p(x, y )}{ p (y,x)}\,.\]
To get \eqref{calvez}  it is enough to observe that  $\sum _{(x,y) \in \cE}  N_{x,y} (\cC_i)\ln \frac{ p(x, y )}{ p (y,x)} = \cA(\cC_i)$. This concludes the proof of our claim \eqref{scappo!}.

 Calling $\hat a_i(t)$ the generalised time--integrated currents associated to the path $\G \left( (Y_s) _{s \in [0,t]}\right)$ and the same cycle basis $\cC_1, \cC_2, \dots, \cC_m$ (use the same definition of $a_i(t)$ referred now to the transformed path), it holds $a_1(t) = \hat a_1(t)$, $a _i (t)= -\hat a_i(t)$ for all $i=2,3,\dots,m$. Combining these identities with  \eqref{scappo!} we get
 \begin{equation*}
 \begin{split}
   e^{- t \cI (\th_1, \th_2, \dots, \th_m) } &  \approx P \left( \frac{a_1 (t)}{t} \sim \th_1, \frac{a_2(t)}{t} \sim \th_2 , \dots, \frac{a_m(t)}{t} \sim \th _m \right)  \\
 &   \approx Q\left( \frac{ a_1 (t)}{t} \sim \th_1, \frac{ a_2(t)}{t} \sim \th_2 , \dots, \frac{  a_m(t)}{t} \sim \th _m \right)
 e^{ t  \sum _{i=2}^m \th_i  \cA( \cC_i)+O(1) } \\
 & = P\left( \frac{ a_1 (t)}{t} \sim \th_1, \frac{ a_2(t)}{t} \sim -\th_2 , \dots, \frac{  a_m(t)}{t} \sim -\th _m \right)
 e^{ t \sum _{i=2}^m  \th_i  \cA( \cC_i)+O(1)} \\
 &= e^{ - t \cI( \th_1, -\th_2, \cdots, -\th_m) + t \sum _{i=2}^m \th_i  \cA( \cC_i)+O(1) }
 \end{split}
 \end{equation*}
From the above approximations, we trivially derive \eqref{ucraina}, thus concluding the proof of Theorem \ref{cicli}.




  \section{Proof of Proposition \ref{suricato}} \label{GC_extra_matrix}
 \medskip
 


Due to the discussion just after Proposition \ref{suricato}, it is enough to exhibit an invertible matrix $U$ satisfying \eqref{patate}.
If $N=2$ it is enough to take  $U= \left (
\begin{matrix}
1 & 0 \\
0 & \frac{\xi_0^+}{\xi_1^-} 
\end{matrix} \right)$ and one can check \eqref{patate} by direct computations. 
\begin{Remark}
In the case $N=2$ one can write explicitly the maximal eigenvalue $\L(\l)$ and check directly \eqref{zoo}. Indeed, it holds
\[ \L(\l) = - \frac{r(0) + r(1)}{2}  + \frac{1}{2} \sqrt{ [r(0)- r(1) ]^2 + 4(\xi_1^+  e^\l + \xi_1^- ) ( \xi_0^+ +  \xi_0^-  e^{-\l}) }  . \]
\end{Remark}

Recalling the definition of the matrix $\cA$ in \eqref{matrixA}  and that   $\D = \ln \frac{ \xi_0^+ \xi_1^+ \cdots \xi_{N-1}^+}{ \xi_0^- \xi_1^- \cdots \xi_{N-1}^-}$  we get
\begin{equation} 
 \cA(-\D-\l) _{i,j}= 
\begin{cases}
-r(i) & \text{ if } i=j\,,\\
\xi_{j} ^+ & \text{ if }0<i \leq N-1,\;\; j=i-1\,,\\
\xi_{j}^- & \text{ if } 0\leq i < N-1, \;\;j=i+1\,,\\
\frac{ \xi_0^+ \xi_1^+ \cdots \xi_{N-1}^+}{  \xi_1^- \xi_2^- \cdots \xi_{N-1}^- } e^\l & \text{ if } i=N-1, j=0\,,\\
\frac{ \xi_0^- \xi_1^-  \cdots \xi_{N-1}^-}{\xi_0^+ \xi_1^+ \dots \xi_{N-2}^+} e^{-\l}   & \text{ if } i=0, j=N-1\,,\\
0 & \text{ otherwise}\,.
\end{cases}
\end{equation}
Hence,  
\[
\cA^{\rm T} (-\D-\l) _{i,j} 
= 
\begin{cases}
-r(i) & \text{ if } i=j\,,\\
\xi_{j-1} ^+ & \text{ if }0<j \leq N-1,\;\; j=i+1\,,\\
\xi_{j+1}^- & \text{ if } 0\leq j < N-1, \;\;j=i-1\,,\\
\frac{ \xi_0^+ \xi_1^+ \cdots \xi_{N-1}^+}{  \xi_1^- \xi_2^- \cdots \xi_{N-1}^- } e^\l & \text{ if } j=N-1, i=0\,,\\
\frac{ \xi_0^- \xi_1^-  \cdots \xi_{N-1}^-}{\xi_0^+ \xi_1^+ \dots \xi_{N-2}^+} e^{-\l}   & \text{ if } j=0, i=N-1\,,\\
0 & \text{ otherwise}\,.

\end{cases}
\]
For example, for $N=3$ we have 
\[\cA (-\D-\l ) = 
	\left( 
	\begin{matrix}
	- r(0) & \xi_1^-  & \xi_2^+ e^{-(\D+\l)}  \\
	 \xi_0^+ & - r(1) & \xi_2^- \\
	 \xi_0^{-} e^{\D+\l} & \xi_1^+ & -r(2)
	\end{matrix} 
	\right),\, \cA ^T(-\D-\l ) = 
	\left( 
	\begin{matrix}
	- r(0) &       \xi_0^+  & \xi_0^{-} e^{\D+\l}  \\
	 \xi_1^-   & - r(1) &  \xi_1^+ \\ 
	 \xi_2^+ e^{-(\D+\l)}  &  \xi_2^- & -r(2)
	\end{matrix} 
	\right)\,.
\]

Take the diagonal matrix $U= (U_{i,j})_{0\leq i,j\leq N-1}$ such that $U_{i,j} = \d_{i,j} c(i)$ and $c(0)=1$, $c(i)= \frac{\xi_0^+ \xi_1^+ \cdots \xi_{i-1}^+}{ \xi_1^- \xi_2^- \cdots \xi_i^-}$ for $1\leq i \leq N-1$.
Then trivially $U^{-1}_{i,j} = \d_{i,j} c(i)^{-1}$ and $[U^{-1} \cA(\l) U]_{i,j}= \cA(\l) _{i,j} \frac{c(j)}{c(i)}$.

Let us  check that the identity $U^{-1} \cA(\l) U= \cA^{\rm T}(-\D-\l)$ holds for each entry.
We have that $[U^{-1} \cA(\l) U]_{i,j}=0$ if we are not in the case $j=i-1,i,i+1$ (with the convention that $0-1=N-1$ and $(N-1)+1=0$). Note that the same holds for $\cA^T(-\D-\l)$. 
Moreover, we have $[U^{-1} \cA(\l) U]_{i,i}= \cA(\l)_{i,i}= -r(i) = \cA^T(-\D-\l)_{i,i} $.  Take now $0< i ,j\leq N-1$. Then 
\begin{align*}
&[U^{-1} \cA(\l) U]_{i,i-1} =  \cA(\l) _{i,i-1} \frac{c(i-1)}{c(i)} = \cA(\l) _{i,i-1}  \frac{   \xi_i^- }{ \xi^+ _{i-1}}=  \xi^+ _{i-1} \frac{   \xi_i^- }{ \xi^+ _{i-1}}=   \xi_i^-= \cA^{\rm T} (-\D-\l) _{i,i-1} \,,
   \\
& [U^{-1} \cA(\l) U]_{j-1,j} =  \cA(\l) _{j-1,j} \frac{c(j)}{c(j-1)} =  \cA(\l) _{j-1,j} \frac{ \xi^+_{j-1} }{\xi_j^-}=
\xi_j^-\frac{ \xi^+_{j-1} }{\xi_j^-}=
 \xi^+_{j-1}=\cA^{\rm T} (-\D-\l) _{j-1,j} 
\end{align*}
Finally, 
\begin{align*}
& [ U^{-1} \cA(\l) U]_{N-1,0 }= \cA(\l) _{N-1,0 }\frac{c (0)}{c(N-1)}= \xi_0^- e^{-\l}\frac
{ \xi_1^- \xi_2^- \cdots \xi_{N-1}^-}{\xi_0^+ \xi_1^+ \cdots \xi_{N-2}^+}=\cA^{\rm T} (-\D-\l) _{N-1,0}\,,
\\
&  [ U^{-1} \cA(\l) U]_{0,N-1}= \cA(\l) _{0,N-1 }\frac{c (N-1)}{c(0)}= \xi_{N-1}^+ e^\l \frac{\xi_0^+ \xi_1^+ \cdots \xi_{N-2}^+}{ \xi_1^- \xi_2^- \cdots \xi_{N-1}^-}=\cA^{\rm T} (-\D-\l) _{0,N-1}\,,
\end{align*}
thus concluding the proof of Proposition \ref{suricato}. 

\section{Technical comments}\label{aiutino}
%
%
%

%
In this appendix we explain how to deal with small fundamental cells in the proof of Theorem \ref{cicli}. 
Let $G=(V,E)$ be a $(\underline{v},\overline{v})$--minimal graph, and let $(z_0 , \ldots , z_n)$ be the unique self--avoiding path from $\underline{v}$ to $\overline{v}$ in $G$. In this section 
we explain how to deal with the cases $n=1,2$ in the proof of Theorem \ref{cicli}, and in particular in the construction of the graph $\tilde{G}$ introduced in Section \ref{GC_extra_cicli}. We treat the case $n=1$ in detail, $n=2$ being analogous (note that in the case $n=2$ the problem with $\tilde G$ would be related with the presence of multiple edges between $\underline{v}$ and $\overline{v}$).

\begin{figure}[!ht]    
\begin{center}
     \centering
  \mbox{\hbox{
  \includegraphics[width=0.8\textwidth]{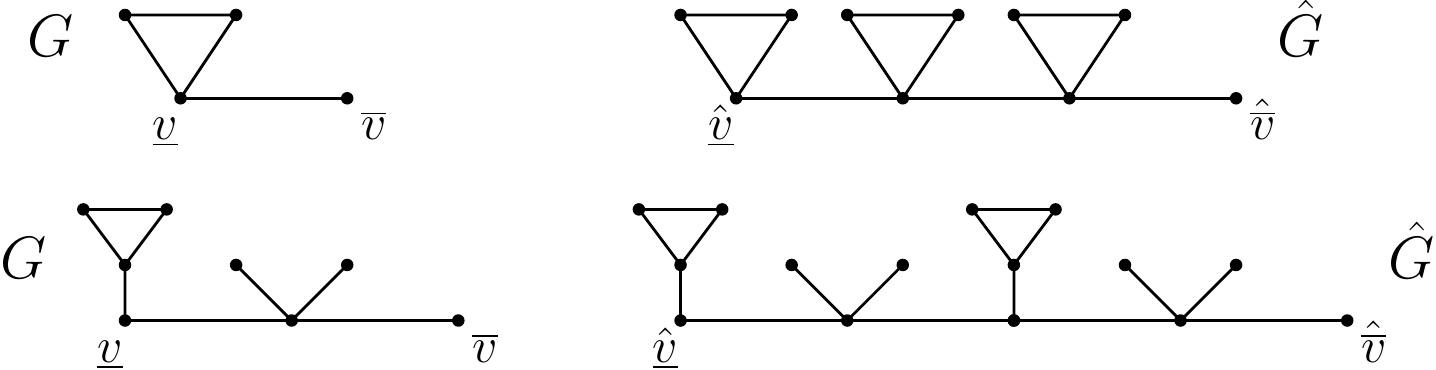}}}
            \end{center}
            \caption{Examples of small fundamental cells $G$ and associated graph $\hat{G}$. For $n=1$ (up), $\hat{G}$ is obtained by gluing $3$ copies of $G$. For $n=2$ (down), it suffices to glue $2$ copies of $G$.}\label{aiutino_fig}
  \end{figure}

When $n=1$, the linear path $(z_0 , \ldots , z_n)$ reduces to the single edge $(\underline{v},\overline{v})$, and the definition of the graph $\tilde{G}$ is not well posed. 
It is therefore useful to \emph{enlarge} the fundamental cell as follows: let $\hat{G}$ be the finite graph obtained by gluing $3$ copies of $G$ so that the vertex $\overline{v} $ of the first (respectively second) copy is identified with the vertex $\underline{v}$ of the second (respectively third) one. Call $\hat{\underline{v}}$ (resp. $\hat{\overline{v}}$) the vertex $\underline{v}$ (resp. $\overline{v}$) of the first (resp. third) copy, as represented in Fig.\ \ref{aiutino_fig}. It is easy to see that if $G$ is $(\underline{v},\overline{v})$--minimal, then $\hat{G}$ is $(\hat{\underline{v}}, \hat{\overline{v}})$--minimal. 
It follows that we can regard $\hat{G}$ as a new fundamental cell, that we use to build a quasi--1d graph $\hat{\cG}$.

Let $X^*_t$, $\hat{X}^*_t$ denote the skeleton processes associated to the CTRW $X_t$ considered as a process on $\cG$, $\hat{\cG}$ respectively.
Then we have 
	\[ | 2 X_t^* - \hat{X}_t^* | \leq 2 \]
for all $t\geq 0$, from which we deduce that the processes $2 X_t^* $ and $ \hat{X}_t^*$ have the same asymptotic properties. In particular, if $\hat{X}_t^* /t $ satisfied a LDP with rate function $I(\th )$, then 
	\[ \bbP \Big( \frac{ X_t^*}{t} \approx \th \Big) 
	\approx
	\bbP \Big( \frac{ \hat{X}_t^*}{t} \approx 2 \th \Big) 
	\approx 
	e^{-t I(2\th )} \, , \]
i.e. $X_t^* /t$ satisfies itself a LDP with rate function $I(2\th )$. 
The study of the large fluctuations and GC symmetry of the process $X^*_t$ can therefore be reduced to the one of the process $\hat{X}^*_t$, with the advantage that the latter is associated to a larger fundamental graph $\hat{G}$.
	
%
%
%
%

\medskip

\medskip

\noindent {\bf Acknowledgements.}   V. Silvestri thanks the Department of Mathematics in University ``La Sapienza''
for the hospitality and   support.  She also   acknowledges the support of the UK Engineering and Physical Sciences Research Council (EPSRC) grant EP/H023348/1 for the University of Cambridge Centre for Doctoral Training, the Cambridge Centre for Analysis.

\end{document}